\documentclass[
  aps,
  prd,
  twocolumn,                 
  preprintnumbers,
  amsmath,
  amssymb,
  nofootinbib,
  longbibliography
]{revtex4-1}
\pdfoutput=1

\usepackage[utf8]{inputenc}
\usepackage[T1]{fontenc}
\usepackage{microtype}        

\usepackage{graphicx}
\usepackage{mathtools}        
\usepackage{amsthm}
\usepackage{bm}               
\usepackage{bbm}              
\usepackage{array}
\usepackage{tabularx}
\usepackage{booktabs}
\usepackage{multirow}
\usepackage{graphicx}
\usepackage{wrapfig}          

\usepackage{xcolor}
\definecolor{nicered}{rgb}{0.7,0.1,0.1}
\definecolor{nicegreen}{rgb}{0.1,0.5,0.1}
\usepackage[colorlinks=true,linkcolor=nicered,citecolor=blue,urlcolor=blue]{hyperref}

\usepackage{enumitem}
\usepackage{verbatim}
\usepackage{tikz}
\usepackage{stackengine}
\usepackage{cancel}
\usepackage[normalem]{ulem}    

\usepackage{setspace}

\numberwithin{equation}{section}

\newcommand{\be}{\begin{equation}}
\newcommand{\ee}{\end{equation}}
\newcommand{\bea}{\begin{aligned}} 
\newcommand{\eea}{\end{aligned}}

\allowdisplaybreaks

\begin{document}
\linespread{1.28}\selectfont

\title{New Isocurvature Constraints from JWST UV Luminosity Function}

\author{Raymond T. Co}
\thanks{rco@iu.edu }
\affiliation{Physics Department, Indiana University, Bloomington, IN 47405, USA}
\author{Sai Chaitanya Tadepalli}
\thanks{saictade@iu.edu (for correspondence) }
\affiliation{Physics Department, Indiana University, Bloomington, IN 47405, USA}

\begin{abstract}
We constrain uncorrelated primordial isocurvature perturbations using a combination of large- and small-scale cosmological probes, with the small-scale data provided by the ultraviolet luminosity function (UVLF)---a measure of number density of galaxies as a function of UV brightness. We consider several isocurvature modes, including cold dark matter, baryon, neutrino density, neutrino velocity, and dark radiation perturbations. The isocurvature power spectrum is modeled using two independent parameterizations: a broken power law and a running power law, without fixing the spectral index a priori. Our analysis combines large-scale data from the Cosmic Microwave Background (CMB), baryon acoustic oscillations, and Type Ia supernovae with small-scale constraints from UVLF measurements obtained by \textit{HST} and \textit{JWST}. The UVLF probes matter fluctuations over a continuous range of intermediate scales, $k \sim 0.5$--$10~\mathrm{Mpc}^{-1}$ over a wide range of redshift $4\lesssim z \lesssim 13$, providing a direct handle on structure formation in a regime where constraints on the scale dependence of isocurvature perturbations remain comparatively limited. Our result represents the first UVLF-based constraint on model-agnostic isocurvature perturbations carried by various components. We construct $68\%$ and $95\%$ credible envelopes in $k$-space for the allowed isocurvature power and find good agreement between the envelopes for the $95\%$ envelope across a wide range of scales, indicating that our constraints are mostly insensitive to the assumed power-law form.

\end{abstract}

\maketitle

\vspace{1cm}

\begingroup
\hypersetup{linkcolor=black}
\renewcommand{\baselinestretch}{1.26}\normalsize
\tableofcontents
\renewcommand{\baselinestretch}{2}\normalsize
\endgroup

\section{Introduction}\label{sec:intro}

The remarkable success of the $\Lambda$CDM framework rests on a simple and powerful assumption: the primordial fluctuations that seeded cosmic structure were predominantly adiabatic. Measurements of the Cosmic Microwave Background (CMB) temperature and polarization anisotropies, together with large-scale structure observations, have confirmed this picture with increasing precision \cite{Planck:2018vyg,AtacamaCosmologyTelescope:2025blo}. Yet the absence of detected isocurvature fluctuations (orthogonal component) remains one of the strongest empirical constraints on physics beyond the minimal inflationary scenario \cite{Planck:2018jri,ACT:2025tim}.

Isocurvature perturbations can arise in a wide range of settings, both during and after inflation. During inflation, additional light degrees of freedom whose fluctuations are not aligned with the inflaton trajectory can generate entropy perturbations, producing variations in the relative number densities of cosmic components rather than in the total energy density. Well-known examples include axion dark matter \cite{Seckel:1985tj}, curvaton models \cite{Enqvist:2001zp,Lyth:2001nq}, moduli fields \cite{Moroi:2001ct}, and more general multi-field inflationary scenarios \cite{Polarski:1994rz,Byrnes:2006fr}. However, isocurvature is not limited to spectator-field fluctuations during inflation. Post-inflationary dynamics can also generate substantial isocurvature power \cite{Chung:1998zb}. In all of these cases, the amplitude and shape of the isocurvature spectrum are tied to the underlying microphysics and to the cosmic epoch at which the relevant field or species is produced, making isocurvature bounds a sensitive probe of both inflationary and post-inflationary physics.

As in the adiabatic case, primordial isocurvature perturbations are most commonly parameterized by simple power-law spectra, in which the variance of the primordial fluctuations scales with wavenumber. Such parameterizations are well motivated and provide an adequate description for a broad class of models. However, they do not exhaust the range of scale dependence expected in realistic scenarios. 

A variety of physically motivated mechanisms, both during and after inflation, can produce isocurvature spectra with nontrivial scale dependence, including broken or featureful shapes rather than a single unbroken power law. These scenarios often involve dynamical or out-of-equilibrium field evolution. Inflationary mechanisms, such as axion-like spectator models with Hubble-induced mass terms—where the Peccei--Quinn radial field evolves out of equilibrium during inflation \cite{Kasuya:2009up,Chung:2021lfg}, or rotating complex scalar fields in the conformal limit \cite{Chung:2024ctx}, can generate a wide range of spectral tilts, including strongly blue spectra with model-dependent scale dependence. By contrast, post-inflationary mechanisms, including symmetry-breaking field fluctuations \cite{Enander:2017ogx}, entropy perturbations generated during phase transitions \cite{Freese:2023fcr,Elor:2023xbz}, and spatial variations in primordial black hole abundance \cite{Afshordi:2003zb}, typically produce spectra that grow as $\propto k^3$ on large scales due to causality, before transitioning at a characteristic scale set by the underlying microphysics. In both cases, the spectrum generally exhibits a turnover or break beyond which the growth is suppressed or saturates. As shown in Ref.~\cite{Chung:2015tha}, such a feature is a generic expectation for large blue-tilted isocurvature spectra from linear spectators and is often required for their observational viability.

More generally, smooth departures from a pure power law can arise from time-dependent dynamics of the isocurvature field during inflation or from transitions in the post-inflationary evolution. For a recent mechanism generating blue-tilted power spectrum with a scale-dependent running over a wide range of scales, see Ref.~\cite{Tadepalli:2026mdc}. Allowing for running provides a convenient phenomenological description of mild, continuous scale dependence without introducing sharp features; in particular, a negative running can mimic a gradual suppression of power on small scales, in contrast to the abrupt transition of a broken power-law spectrum. These considerations motivate extending the standard power-law framework to a broader class of parameterizations.

On the observational side, recent measurements from \textit{JWST} at high redshifts, together with existing \textit{HST} data and updated UVLF likelihoods that incorporate astrophysical marginalization over nuisance parameters \cite{Sabti:2021xvh}, make it possible to combine updated CMB constraints with small-scale structure information probing approximately $0.5\,{\rm Mpc}^{-1} \lesssim k \lesssim 10\,{\rm Mpc}^{-1}$. Since this range begins near the scales where the constraining power of the primary CMB declines, UVLF measurements provide a natural extension of the observational lever arm into the small-scale regime. The combination of CMB and UVLF data therefore makes it possible to place comparatively smooth bounds on the allowed isocurvature amplitude across the transition from CMB scales to genuinely small scales, including the range where spectra with a break, running, or other nontrivial scale dependence may depart from a simple power law. 

Other probes of small-scale structure, such as the Lyman-$\alpha$ forest and 21-cm line observations, can in principle provide complementary constraints on similar or even smaller scales. Recent 21-cm line measurements have set increasingly stringent upper limits on the 21-cm power spectrum \cite{HERA:2022wmy,deKruijf:2024voc}. However, these limits remain largely consistent with thermal noise and are primarily sensitive to astrophysical processes such as intergalactic medium heating and reionization. Lyman-$\alpha$ forest analyses have recently reached a level of precision to place highly competitive, and in some cases stronger, constraints on small-scale power for specific model assumptions~\cite{Pavicevic:2025gqi,Garcia-Gallego:2026phh}. However, these analyses rely on detailed hydrodynamical simulations to accurately model the intergalactic medium and its thermal history, making them computationally intensive and less straightforward to generalize across a wide range of primordial power spectra and initial conditions. By contrast, UVLF measurements provide a more direct and flexible probe of small-scale structure across a wide range of comoving scales and redshifts, and are significantly easier to implement. This makes them especially well suited for systematically exploring a broad class of scale-dependent isocurvature spectra.

In this work, we combine CMB likelihoods with recent UVLF measurements to derive new bounds on primordial isocurvature perturbations. 

The UVLF dataset has been utilized recently in Refs.~\cite{Urrutia:2025fvp,Gorghetto:2025uls} to probe white-noise type contributions that exhibit a characteristic $k^3$ scaling of the power spectrum amplitude as a function of comoving scales, such as those generated in post-inflationary axion scenarios. 

Our goal is to present constraints on the allowed isocurvature amplitude over a broad range of scales up to $k \lesssim 10~\rm{Mpc}^{-1}$ for several isocurvature carriers, including cold dark matter, baryons, neutrino density, neutrino velocity, and dark radiation, while maintaining a model-agnostic approach by treating the spectral tilt as a free parameter and exploring two distinct parameterizations of the power spectrum shape.

The remainder of this paper is organized as follows. In Sec.~\ref{Sec:theory} we introduce standard definitions of the primordial adiabatic and isocurvature perturbations as applied within linear Boltzmann solvers. We further parametrize the spectral shapes that we study in this work. Sec.~\ref{Sec:dataset} details our datasets, Bayesian MCMC setup, including the likelihood components and baseline cosmological assumptions. Our main results are presented in Sec.~\ref{Sec:Results}, followed by a summary and discussion in Sec.~\ref{Sec:discussion}. In App.~\ref{App:uvlf_detail} we describe the implementation of the UVLF observable, its associated nuisance parameters and integration into our cosmological inference pipeline. In App.~\ref{app:spectrum_reconstruction} we discuss the envelope reconstruction procedure implemented in this work.

\section{Theory and Parameterization}\label{Sec:theory}

\subsection{Primordial Adiabatic and Isocurvature Perturbations}

In the presence of multiple light degrees of freedom during inflation, the primordial perturbations may contain both adiabatic and isocurvature components. The adiabatic mode corresponds to fluctuations in the total energy density with fixed relative composition, such that the relative entropy perturbations between species vanish. Isocurvature perturbations, by contrast, describe fluctuations in the relative abundances or velocities of different species, with vanishing initial total curvature perturbation. These may arise either in the density sector, as relative number-density perturbations at fixed total energy density, or in the velocity sector, as relative velocity perturbations between different fluids.

To determine the leading order evolution of cosmological perturbations in various fluids such as photons, dark matter, baryons, and neutrinos, one must solve the coupled linearized Einstein and fluid equations, initialized with the adiabatic and isocurvature primordial fluctuations. At the level of linear perturbations, the total scalar dimensionless primordial power spectrum is given by the sum of uncorrelated adiabatic and isocurvature components,
\begin{equation}
\Delta^2_{\mathrm{s}}(k)=\Delta^2_{\mathrm{ad}}(k)+\Delta^2_{\mathrm{iso}}(k).
\end{equation}
The dimensionless power spectrum as given above is defined as
\begin{equation}
\Delta^2(k)\equiv \frac{k^3}{2\pi^2}P(k),
\end{equation}
where the usual power spectrum $P(k)$ is defined through the Fourier-space two-point function,
\begin{equation}
\langle \delta(\mathbf{k})\,\delta(\mathbf{p})\rangle
=(2\pi)^3\delta^{(3)}(\mathbf{k}+\mathbf{p})\,P(k).
\end{equation}
Under the assumption that the adiabatic and isocurvature modes are uncorrelated, their contributions simply add to give the total primordial scalar spectrum.

We parametrize the adiabatic spectrum in the standard power-law form,
\begin{equation}
\Delta^2_{\rm ad}(k)
=
A_{\rm ad}
\left( \frac{k}{k_p} \right)^{n_{\rm ad} - 1},
\end{equation}
with an amplitude $A_{\rm ad}$ and a scalar spectral index $n_{\rm ad}$ defined at the pivot scale $k_p=0.05$/Mpc.\footnote{Within the literature, $A_{\rm ad}$ and $n_{\rm ad}$ are often denoted by the symbols $A_{\rm s}$ and $n_{\rm s}$ respectively.} Current CMB measurements~\cite{Planck:2018jri} indicate that the primordial fluctuations are predominantly adiabatic, with 
\[
A_{\rm ad}\simeq 2.1\times 10^{-9}, \qquad n_{\rm ad}\simeq 0.97.
\]
No significant detection of isocurvature has been made, although weak hints were reported for uncorrelated blue-tilted CDI models in Ref.~\cite{Chung:2017uzc}.

\subsection{Broken Power-Law Isocurvature Spectrum}

To allow for scale-dependent structure, we consider a broken power-law parameterization for the isocurvature spectrum,
\begin{equation}
\Delta^2_{\rm iso}(k)
=
A_{\rm iso}
\begin{cases}
\left( \dfrac{k}{k_p} \right)^{n_{\rm iso}-1}, & k < k_c, \\
\left( \dfrac{k_c}{k_p} \right)^{n_{\rm iso}-1}, & k \ge k_c~.
\end{cases}\label{eq:broken-power-law}
\end{equation}
Here:
\begin{itemize}
\item $A_{\rm iso} \equiv \Delta^2_{\rm iso}(k_p)$ is the isocurvature amplitude at the pivot scale $k_p$,
\item $n_{\rm iso}$ is the spectral index for modes below the break,
\item $k_c$ is the break scale.
\end{itemize}

Above the break scale \(k_c\), we assume for simplicity that the isocurvature spectrum saturates to a scale-independent amplitude, rather than continuing as a pure power law. This broken power-law form is meant to capture scenarios in which the isocurvature sector develops scale dependence only over a limited range of scales, with the evolution effectively changing regime at a characteristic scale. Such behavior can be motivated either by causal post-inflationary physics or by time-dependent dynamics during inflation ~\cite{Kasuya:2009up,Chung:2021lfg,Chung:2024ctx,Enander:2017ogx,Freese:2023fcr,Elor:2023xbz,Afshordi:2003zb,Tadepalli:2026mdc}. 

\subsection{Running Isocurvature Spectrum}

As a complementary scenario, we consider an isocurvature spectrum without an explicit cutoff but allowing for scale-dependent running of the spectral index. In this case the standard running spectrum is defined as
\begin{equation}
\Delta^2_{\rm iso}(k)
=
A_{\rm iso}
\left( \frac{k}{k_p} \right)^{n_{\rm iso}(k)-1},\label{eq:powerlaw_running}
\end{equation}
where the scale-dependent spectral index is given by
\begin{equation}
\frac{d\ln \Delta^2_{\rm iso}(k)}{d\ln k} + 1=n_{\rm iso}(k)
=
n_{\rm iso}
+
\alpha_{\rm iso}
\ln\left( \frac{k}{k_p} \right).\label{eq:scale-dependent-niso}
\end{equation}

Equivalently, the spectrum can be written as
\begin{equation}
\Delta^2_{\rm iso}(k)
=
A_{\rm iso}
\left( \frac{k}{k_p} \right)^{n_{\rm iso}-1
+
\frac{1}{2}\alpha_{\rm iso}\ln(k/k_p)}.
\end{equation}

Here:
\begin{itemize}
\item $n_{\rm iso}$ is the spectral index at the pivot scale,
\item $\alpha_{\rm iso} \equiv d n_{\rm iso}(k)/d \ln k$ is the running of the isocurvature spectral index.
\end{itemize}

This parameterization captures smooth deviations from a pure power law and allows for enhanced or suppressed small-scale power without introducing a sharp feature. In particular, a negative running can mimic gradual turnovers or extended suppressions in the spectrum, providing a more flexible description than a broken power law, which instead imposes a localized change in slope.

In summary, we consider two classes of isocurvature models with free parameters as follows:
\begin{itemize}
\item Broken power-law model:
\(
\{ A_{\rm iso},\, n_{\rm iso},\, k_c \}.
\)
\item Running model:
\(
\{ A_{\rm iso},\, n_{\rm iso},\, \alpha_{\rm iso} \}.
\)
\end{itemize}

In both cases, the total primordial spectrum is given by the sum of uncorrelated adiabatic and isocurvature contributions, and cosmological observables are computed by evolving the full set of linear perturbation equations with these initial conditions.

\subsection{Isocurvature Modes: CDI, BI, NDI, NVI and DRDI}

We consider five classes of primordial isocurvature initial conditions: cold dark matter density isocurvature (CDI), baryon density isocurvature (BI), neutrino density isocurvature (NDI), neutrino velocity isocurvature (NVI), and dark radiation density isocurvature (DRDI). 
For density isocurvature modes, a gauge-invariant entropy perturbation between two species $i$ and $j$ may be defined~as
\begin{equation}
S_{ij} \equiv 3(\zeta_i-\zeta_j),
\end{equation}
where $\zeta_i$ denotes the curvature perturbation on hypersurfaces of uniform energy density for species $i$. This definition applies to CDI, BI, NDI and DRDI. For example, the CDM isocurvature perturbation may be written~as
\begin{equation}
S_{c\gamma}
=
\frac{\delta \rho_c}{\rho_c}
-\frac{3}{4}\frac{\delta \rho_\gamma}{\rho_\gamma},
\end{equation}
on superhorizon scales.  Velocity isocurvature modes are instead characterized by gauge-invariant relative velocity perturbations. Under a scalar gauge transformation, the velocity potential of each species shifts by the same amount, so the difference
\begin{equation}
V_{ij}\equiv v_i-v_j
\end{equation}
is gauge invariant. The neutrino velocity isocurvature mode is therefore described by a nonzero initial relative velocity $V_{\nu\gamma}$. While distinct from the standard NVI initial condition considered here, recent work on kinetic isocurvature perturbations provides a useful example of how nonstandard velocity- or momentum-type isocurvature perturbations can arise in physically motivated scenarios~\cite{Bae:2026hly}.

For the full set of superhorizon initial conditions for these isocurvature modes, as implemented in linear Boltzmann solvers, we refer the reader to Ref.~\cite{Bucher:1999re}.

CDI is well motivated in axion and spectator-field scenarios and is therefore taken as our primary matter isocurvature mode. CMB signatures from BI are nearly degenerate with those of CDI, differing mainly by an overall normalization set by the background matter fractions. In particular, the total matter entropy perturbation satisfies
\begin{equation}
S_m=
\frac{\Omega_c}{\Omega_m}S_c
+
\frac{\Omega_b}{\Omega_m}S_b,
\qquad
\Omega_m=\Omega_c+\Omega_b,
\end{equation}
so that constraints on BI can be inferred directly from those on CDI through the corresponding density rescaling. Thus we do not treat BI as an independently sampled mode.

The neutrino modes must instead be analyzed separately. The NDI mode corresponds to a nonzero neutrino-photon entropy perturbation in the radiation sector, while the NVI mode corresponds to a nonzero initial neutrino-photon relative velocity. Both generate distinct initial conditions for the subsequent perturbation evolution and lead to characteristic signatures in the CMB anisotropies that cannot be mapped onto CDI by a simple rescaling.

In contrast, the DRDI mode is closely analogous to NDI~\cite{Ghosh:2021axu}. If one defines the density fractions as $R_i = \bar{\rho}_i/\bar{\rho}_r$ where $r$ stands for overall radiation, then in the limit of a small dark-radiation fraction ($R_{\rm dr}\ll 1$), its superhorizon initial conditions reduce to those of NDI under the replacement $R_{\nu}\rightarrow R_{\rm dr}$, leading to an overall normalization difference set by the radiation fractions.\footnote{This mapping holds only for small $R_{\rm dr}$, where the dark-radiation component does not significantly modify the background expansion or total radiation content, so that DRDI reduces to an NDI-like mode up to an overall normalization.} It follows that 
\[R_{\rm dr}^2 A_{\rm DRDI} \approx \left(\frac{R_\nu}{1-R_\nu}\right)^2 A_{\rm NDI}.   \]As a result, constraints on DRDI can be obtained from those on NDI by an appropriate rescaling given by
\[
R_{\rm dr}^2 A_{\rm DRDI} \approx 0.48 A_{\rm NDI},
\] where we take $R_\nu \approx 0.41$ for the standard neutrinos. Thus, similar to BI we do not treat DRDI as an independent mode, but infer its limits from those of NDI.

Accordingly, we perform dedicated analyses for CDI, NDI, and NVI, while BI and DRDI constraints are obtained from the CDI and NDI results respectively through density-weighted rescaling as shown above. For CDI and NDI, we consider both broken power-law and running isocurvature models. For NVI, we restrict our analysis to the running parameterization for simplicity. We note that, as shown in Sec.~\ref{Sec:Results}, both parameterizations lead to similar upper limits on the primordial isocurvature amplitude for CDI/BI and NDI across the broad range of scales, $0.001 \lesssim k \lesssim 5\,{\rm Mpc}^{-1}$ .

\section{Datasets and Analysis Pipeline}\label{Sec:dataset}

\subsection{Cosmological Datasets}

Our constraints on primordial isocurvature modes combine information from multiple cosmological probes that span a broad range of physical scales. We incorporate the following datasets with updated/recent likelihoods in most cases:

\paragraph{CMB Anisotropies and Lensing.}
For the primary CMB anisotropies we use a hybrid Planck likelihood: at low multipoles we adopt the Planck 2018 low-$\ell$ temperature likelihood together with the \textsf{SRoll2} low-$\ell$ polarization likelihood \cite{Pagano:2019tci}, while at high multipoles we use the \textsf{CamSpec} TT/TE/EE likelihood constructed from the Planck PR4 (\textsf{NPIPE}) maps, including data up to $\ell_{\max}=2500$ with a free calibration $A_{\rm Planck}$~\cite{Planck:2018lkk, Efstathiou:2019mdh,Rosenberg:2022smo}. In the high-$\ell$ analysis we vary the standard \textsf{CamSpec} foreground, beam, and calibration nuisance parameters jointly with the cosmological parameters and marginalize over them in the posterior. For CMB lensing we use the ACT DR6 lensing likelihood with the ACT-Planck combined \texttt{actplanck\_baseline} variant (up to $L_{\max}=4000$), which constrains the lensing potential power spectrum and tightens constraints on the integrated late-time clustering~\cite{Carron:2022eyg,ACT:2023dou,ACT:2023kun}.

\paragraph{Baryon Acoustic Oscillations (BAO).}
Baryon acoustic oscillation measurements from spectroscopic surveys are incorporated to anchor the expansion history and distance scale. We include the latest high-precision DR2 BAO measurements from galaxy and quasar samples released by the Dark Energy Spectroscopic Instrument (DESI), which provide distance constraints across multiple redshift bins. These BAO data constrain cosmological expansion independently of the primordial power spectrum shape and are crucial for robust parameter inference~\cite{DESI:2024uvr,DESI:2024lzq,DESI:2024mwx}.

\paragraph{Type Ia Supernovae and Expansion History.}
We incorporate standardized luminosity distance measurements from Type Ia supernovae compilations (e.g., Pantheon+), which constrain the low-redshift expansion history \cite{Pan-STARRS1:2017jku,Jones:2017udy}. These data complement the BAO measurements and ensure that late-time cosmological parameters are consistently determined across the full redshift range.

\paragraph{Ultraviolet Luminosity Function (UVLF)}
In addition to large-scale data, we employ measurements of the ultraviolet luminosity function of high-redshift galaxies to probe small-scale structure. Ultraviolet luminosity function (UVLF) data from deep imaging surveys, in particular those collected with the \textit{Hubble Space Telescope} (\textit{HST}) at $z\simeq 4\,\text{--}\,8$ \cite{Oesch_2018,Bouwens_2021} and the \textit{James Webb Space Telescope} (\textit{JWST}) at $z\simeq 9\,\text{--}\,13$~\cite{donnan2024jwst}, spanning magnitudes from the bright exponential cutoff ($M_{\rm UV}\sim -22$) to the faint-end power-law tail ($M_{\rm UV}\sim -16$) over redshifts $4 \lesssim z \lesssim 13$. We do not include \textit{JWST} samples from even higher redshifts as there are currently no robust spectroscopically confirmed galaxy samples at $z>14.5$. To account for the cosmic
variance and the theoretical uncertainty in analytic halo mass function (HMF) formula (see Fig.~3 in Ref.~\cite{Co:2025hbi}) we adopt a conservative $20\%$ floor on the error of each data point. This in turn makes our results slightly conservative. The UVLF measurements from high redshifts extend the dynamic range of our analysis to comoving wavenumbers $k \sim \mathcal{O}(10)\,\mathrm{Mpc}^{-1}$, where isocurvature and new physics signatures can affect galaxy formation and abundance. By comparing model predictions for the halo mass function and UV luminosity distribution to the observed UVLF, we incorporate small-scale information that is otherwise inaccessible to traditional large-scale probes. For details regarding the astrophysical and star-formation modeling, UVLF nuisance parameters, UVLF data source, halo mass function parameterization etc., we refer the readers to App.~\ref{App:uvlf_detail} and Sec.~3 of Ref.~\cite{Co:2025hbi}.

\subsection{Sampler and Pipeline}
We perform Markov-chain Monte-Carlo (MCMC) sampling using \textsf{Cobaya} \cite{Torrado:2020dgo}, fitting jointly to CMB, BAO, Type Ia supernovae (SNe), and UVLF likelihoods. Since the original UVLF likelihood, GALLUMI, was developed for \textsf{MontePython}~\cite{brinckmann_montepython_2019}, we adapted and validated its implementation within \textsf{Cobaya} and is publicly accessible at \url{https://github.com/tsaic2808/UVLF_likelihood_cobaya}. Each MCMC run consists of four independent parallel chains initialized from a pre-computed covariance matrix. Final constraints are derived from the converged chains using \textsf{GetDist}~\cite{lewis_getdist_2025}.

\subsection{Cosmology and Sampling}
We consider a spatially flat $\Lambda$CDM model for our fiducial cosmology with massless neutrinos and a cosmological constant $\Lambda$. This model includes six free parameters \{$\omega_{b},~\omega_{\rm cdm},~\tau_{\rm reio},~\theta_{s,100},~A_{\rm ad},~n_{\rm ad}$\} that we sample within our MCMC analysis. When considering neutrino isocurvature fluctuations (NDI/NVI), we take a single species of massive neutrino with mass $60~$meV. All species are assumed to carry nearly scale-invariant adiabatic perturbations. In each MCMC analysis, we consider that only one species carries the uncorrelated isocurvature initial conditions. We evaluate CMB anisotropy and matter power spectra using the linear Boltzmann solver \textsf{CLASS} \cite{Blas:2011rf}.

Similar to the strategy of previous broken power-law analyses (e.g. Ref.~\cite{Bae:2026hly}), when parameterizing the isocurvature spectrum with a
broken power law we consider the break scale $k_c$ as a fixed parameter.\footnote{We modified \textsf{CLASS} code to allow a break/cutoff in the isocurvature power spectrum.} In that work, the authors fixed the large-scale spectral index to a specific value
($n_{\rm iso}=4$) for all isocurvature modes and directly constrained the
isocurvature amplitude at each chosen break scale, $\mathcal{A}(k_c)$. While this
choice yields useful constraints within a specific phenomenological template, it could
limit the analysis to a narrow subclass of possible scale dependences.

In contrast, our analysis does not impose a prior fixing $n_{\rm iso}$. Instead,
we allow the scale dependence of the isocurvature spectrum to vary and assess
the extent to which it can be constrained by the combined large- and small-scale datasets primarily from CMB and UVLF. To do so in a way that reduces prior-volume effects
and anchors the inference at physically relevant scales, we explicitly sample
the primordial dimensionless isocurvature amplitudes, $\mathcal{A}(k)\equiv\Delta^2(k)$, at two widely separated comoving wavenumbers
\begin{equation}
\mathcal{A}_{\rm iso}(k_1),\quad k_1 = 0.05\ \mathrm{Mpc}^{-1},
\end{equation}
and
\begin{equation}
\mathcal{A}_{\rm iso}(k_2),\quad k_2 = 3\ \mathrm{Mpc}^{-1},
\end{equation}
which correspond to the scales most relevant to the large-scale CMB and the
small-scale UVLF, respectively. Anchoring the isocurvature amplitude at these
two pivot points provides a flexible description of the scale dependence
without fixing the tilt a priori; the degree to which that tilt is actually
constrained is then left to the data. This approach is similar to that adopted by several previous CMB-based analyses such as Planck~\cite{Planck:2018jri} and ACT~\cite{ACT:2025tim}.

For a given choice of break scale $k_c$, we perform an MCMC run by sampling the pair 
\[
\{\mathcal{A}_{\rm iso}(k_1),\,\mathcal{A}_{\rm iso}(k_2)\}
\]
over broad, uniform priors, together with the standard cosmological and nuisance parameters. Under the
broken power-law model, the amplitudes at $k_1$ and $k_2$ uniquely determine
the amplitude at the break,
\( \mathcal{A}_{\rm iso}(k_c),
\)
through the spectral scaling in the two regimes. We compute
$\mathcal{A}_{\rm iso}(k_c)$ as a derived parameter in each chain and use the
posterior distribution of this derived quantity to obtain 95\% credible
intervals on the isocurvature amplitude at the break scale. 

This two-pivot sampling approach avoids imposing an arbitrary fixed spectral
index and reduces sensitivity to prior-volume effects. It allows for a more
flexible description of the isocurvature spectrum, so that the inferred limits
on $\mathcal{A}_{\rm iso}(k_c)$ depend as directly as possible on the data over
the scales probed by the combined datasets, rather than on restrictive
assumptions about the spectral shape. By repeating this procedure for multiple
fixed values of $k_c$, we derive upper limit constraints on the broken power-law isocurvature amplitude across a range of relevant scales.

For the running isocurvature spectrum, we adopt a related but simpler sampling strategy. Within \textsf{CLASS}, primordial isocurvature with a running is specified by its amplitude at a pivot scale \(k_p\),
\begin{equation}
    f_{\rm iso} = \sqrt{A_{\rm iso}/A_{\rm ad}}\label{eq:fiso}
\end{equation}together with a spectral index \(n_{\rm iso}\) and a running $\alpha_{\rm iso}$ as defined in Eqs.~(\ref{eq:powerlaw_running}) and (\ref{eq:scale-dependent-niso}). As in the broken power-law case, we sample the amplitudes $\mathcal{A}_{\rm iso}(k_1)$ and $\mathcal{A}_{\rm iso}(k_2)$ along with $n_{\rm iso}$, and infer $\alpha_{\rm iso}$ from the assumed functional form. 

Unlike the broken power-law analysis, which requires multiple MCMC runs over a range of fixed break scales $k_c$, the running parameterization does not introduce an additional scale. Instead, the spectrum is fully specified by a smooth functional form with global correlations across wavenumbers. The allowed spectra form a tightly correlated family, and a limited set of runs is adequate to map the posterior support across the relevant range of scales. As a result, varying the choice of anchor scales primarily affects how efficiently different regions of $k$-space are explored, rather than defining distinct model families. In practice, we find that performing two MCMC analyses, differing only in the choice of the lower anchor scale, $k_1=0.01$~Mpc$^{-1}$ and $k_1=0.05$~Mpc$^{-1}$, is sufficient to capture the range of isocurvature spectra consistent with the data.

For the UVLF nuisance parameters, we assume all of them to be redshift independent, except for the slope of the star-formation efficiency at the faint-end of the UV magnitude $\alpha_\star$ as defined in App.~\ref{App:uvlf_detail}, for which we assume an additional linear slope parameter $\alpha_{\star,s}$. This simplifying choice has been shown to fit the high-redshift \textit{JWST} dataset better with minimal extension of otherwise redshift-independent parameterization \cite{Shadab_uvlf}. It also 
reduces the astrophysical freedom in the galaxy–halo connection and therefore yields a conservative lower limit on the combined low- and high-redshift UVLF data, jointly fitting both \textit{HST} and \textit{JWST} datasets with minimal set of astrophysical nuisance parameters \cite{Shadab_uvlf}.\footnote{Allowing all of the parameters to vary with redshift would introduce additional degeneracies, thereby weakening the resulting constraints.} The priors for the UVLF astrophysical parameters are given in App.~\ref{App:uvlf_detail}.

\section{Results}\label{Sec:Results}
We now present the results on the upper limits for the various isocurvature modes, constrained jointly by the CMB+BAO+SNe+UVLF likelihoods. 

\subsection{Reconstruction of Isocurvature Spectral Envelopes}
For the broken power-law parameterization, rather than presenting constraints on $\mathcal{A}_{\rm iso}(k_c)$ for each assumed break scale $k_c$, we summarize the results directly in terms of an envelope of allowed isocurvature spectra, defined as pointwise upper credible bounds across the ensemble of reconstructed spectra. 

To construct this envelope, we combine posterior samples from the full set of
fixed-$k_c$ MCMC analyses. For each chain, we draw a representative subset of
posterior samples and reconstruct the corresponding spectrum
$\Delta^2_{\rm iso}(k)$ on a common logarithmically spaced $k$ grid. Each
realization is built from the sampled amplitudes at the two anchor scales,
using a power-law form below the break and a constant plateau above it. The
ensemble of reconstructed spectra from all sampled values of $k_c$ is then used
to compute pointwise posterior quantiles at each wavenumber, from which we
derive the $68\%$ and $95\%$ credible upper envelopes. We refer the readers to App.~\ref{app:spectrum_reconstruction} for more details about this reconstruction procedure implemented in our work.

We adopt an analogous reconstruction procedure for the running isocurvature
model. In this case, posterior samples are drawn from the MCMC chain in the
space of the amplitude, tilt, and running parameters, and for each sample we
reconstruct the corresponding spectrum on the same common $k$ grid. The
resulting ensemble is again converted into pointwise $68\%$ and $95\%$
credible envelopes. Presenting both parameterizations in this common
envelope-based form facilitates a direct comparison of the model-dependency on the allowed scale-dependent isocurvature power across the full range of scales probed by
the combined datasets.

By construction, these envelopes should be interpreted as pointwise summaries
of the isocurvature power allowed by the data within the assumed model class,
rather than as the posterior for a single underlying spectral realization. In the broken power-law case, this construction effectively traces a near-maximal envelope of allowed spectra: because the break scale $k_c$ is varied across the ensemble, different wavenumbers can be constrained independently by different subsets of models, allowing the envelope at each $k$ to be saturated by spectra with distinct break locations and amplitudes. 

We emphasize, however, that this construction does not correspond to a non-parametric or “free” reconstruction of the spectrum. Unlike approaches based on binned or independently varying amplitudes at each $k$, the spectra entering the envelope here are restricted to the broken power-law family, within which the power at different scales remains strongly correlated for any given model. The apparent scale-by-scale freedom arises only through the combination of different models with varying $k_c$, rather than from independent degrees of freedom in the spectrum itself.

By contrast, the running parameterization imposes a stronger global correlation across scales, since the spectrum is fully determined by a small set of parameters (amplitude, tilt, and running). As a result, increasing the power on one range of scales generically induces correlated changes elsewhere. For example, a large amplitude at high $k$ typically requires a steeper tilt and/or running, which in turn suppresses power on larger scales or enhances it beyond the pivot in a constrained way. Consequently, the envelope derived from the running model reflects a more restricted subset of spectral shapes and should be viewed as a highly model-dependent projection of the allowed power, rather than a scale-by-scale upper bound in the same sense as the broken power-law case.

Lastly, we note that this construction differs from the commonly adopted approach in which one quotes upper limits on the amplitude at the break scale, $\mathcal{A}_{\rm iso}(k_c)$, separately for each fixed $k_c$. Because the spectral shape correlates power across scales, maximizing the amplitude at a given break scale does not, in general, maximize the power at other wavenumbers. As a result, the commonly plotted constraints on $\mathcal{A}_{\rm iso}(k_c)$ as a function of $k_c$ and the pointwise spectral envelope constructed here represent distinct projections of the same underlying constraints, with the former characterizing parameter limits within individual models and the latter summarizing the allowed power at each scale across the model family.
The latter thus provides a complementary characterization of the scale-dependent constraints implied by the data, expressed directly in $k$-space across the model family. 

An additional advantage of this envelope-based representation is that it can be applied uniformly across different parameterizations. In particular, for the running model, where there is no discrete break scale, the same procedure yields a pointwise spectral envelope directly from a single MCMC analysis, facilitating a direct comparison with the broken power-law case in a common $k$-space representation.

\subsection{Constraints on Isocurvature Spectra}
\begin{figure*}[t]
    \centering
    \includegraphics[width=0.48\textwidth]{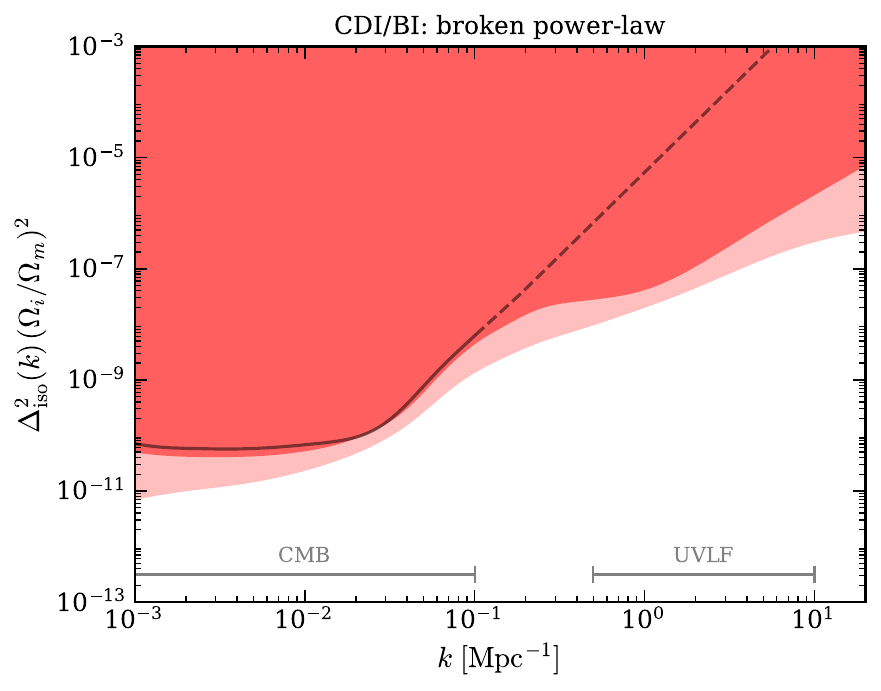}~~
    \includegraphics[width=0.48\textwidth]{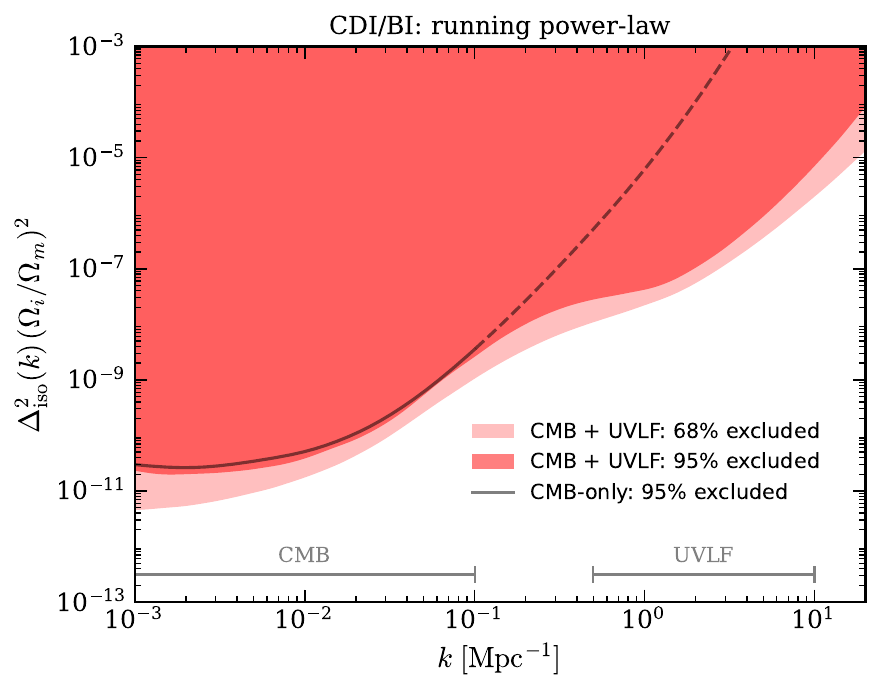}\\[0.3cm]
    \includegraphics[width=0.48\textwidth]{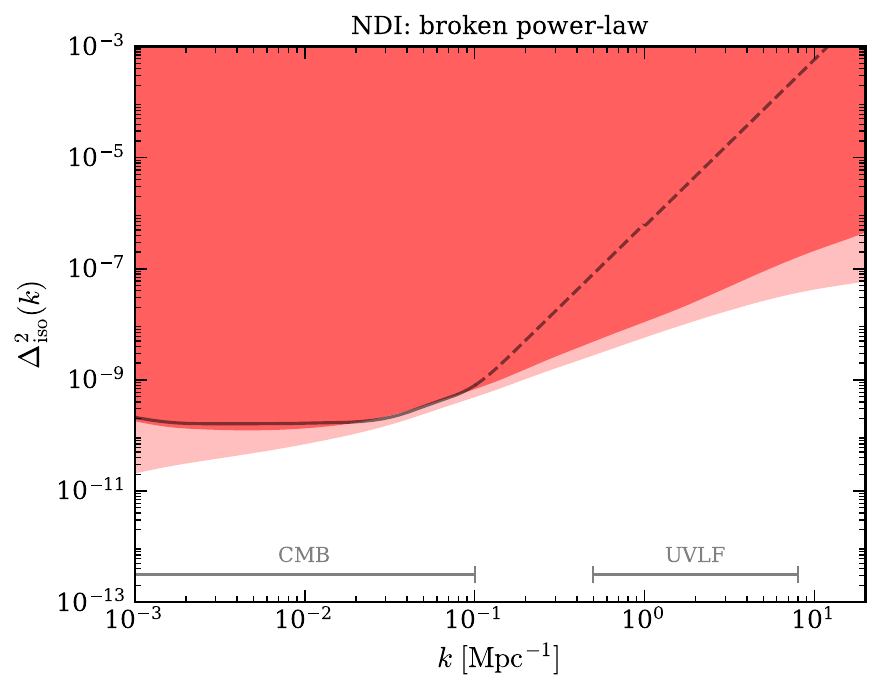}~~
    \includegraphics[width=0.48\textwidth]{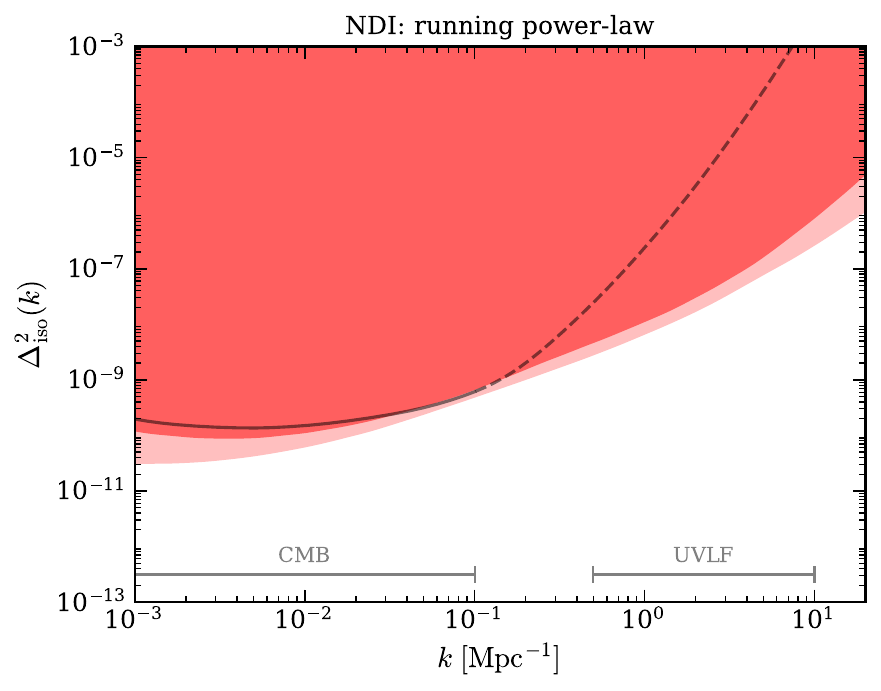}
    \caption{$68\%$ and $95\%$ credible upper envelopes on the primordial
isocurvature power spectrum $\Delta^2_{\rm iso}(k)$ inferred from the combined
CMB and UVLF analysis. The shaded regions indicate the corresponding excluded
parameter space above these envelopes, while the unshaded region denotes the
allowed parameter space. The top row shows the CDI/BI mode, while the bottom row
shows the NDI mode. In each row, the left and right panels correspond to the
broken power-law and running parameterizations, respectively. To highlight the improvement from the inclusion of the UVLF dataset, we show the CMB-only upper bound as a solid gray curve for $k<0.1\,\mathrm{Mpc}^{-1}$ and as a dashed gray curve for $k>0.1\,\mathrm{Mpc}^{-1}$, where the latter represents an extrapolation. Horizontal indicators mark the specific ranges where CMB and UVLF data are most sensitive. }
    \label{fig:isocurvature_envelopes}
\end{figure*}

Figure~\ref{fig:isocurvature_envelopes} shows the $68\%$ and $95\%$ credible upper envelopes of the primordial density isocurvature spectra for the CDI/BI and NDI modes, considering both the broken power-law and running power-law parameterizations. The shaded regions correspond to amplitudes excluded at the indicated credibility levels. The gray solid curve shows the CMB-only upper bound for $k<0.1~{\rm Mpc}^{-1}$, which marks the range of scales over which the CMB directly constrains the isocurvature amplitude.

For the broken power-law parameterization, the spectrum is defined only over the CMB-constrained range and is not extrapolated to smaller scales as a physical constraint. We therefore show, as a gray dashed curve, an illustrative $k^3$ continuation corresponding to the steepest allowed blue tilt by causality, $n_{\rm iso}=4$. By contrast, the running power-law parameterization is a continuous model whose sampled parameters can be evaluated beyond the CMB-sensitive scales. For this case, the small-scale extension shown in the figure is obtained directly from the CMB-only posterior samples.\footnote{In the CMB-only analysis for the running power-law model, we shifted the two anchor scales at $k_{1(2)}$ to $0.002~(0.1)$~Mpc$^{-1}$ so that they adequately cover the CMB range.} Notably, the inclusion of UVLF data improves upon these CMB-only constraints by approximately two to three orders of magnitude at smaller scales.

We find that the broken power-law and running
parameterizations yield very similar upper envelopes over a broad
range of scales $0.001 \lesssim k \lesssim 5\,{\rm Mpc}^{-1}$, where the
combined CMB and UVLF data provide complementary constraints. 
On smaller scales ($k \gtrsim 5\,{\rm Mpc}^{-1}$), the inferred envelopes can exhibit a mild dependence on the assumed spectral parameterization. The resulting limits on $\Delta^2_{\rm iso}(k)$ are therefore robust within the
intermediate range probed jointly by both datasets, but retain some model
dependence outside this region.

The residual differences between the two parameterizations reflect how the
running power-law model redistributes power across widely separated scales. Because the spectrum is constrained by a single amplitude, tilt, and running, a
larger amplitude at small scales ($k \sim k_2$) typically
requires a steeper tilt and/or positive running. This correlated adjustment
enhances the power on higher scales $k \gtrsim k_2$. As a result, the running parameterization tends to yield a higher envelope at small scales
compared to the broken power-law case.

\begin{figure}[t]
    \centering
    \includegraphics[width=0.48\textwidth]{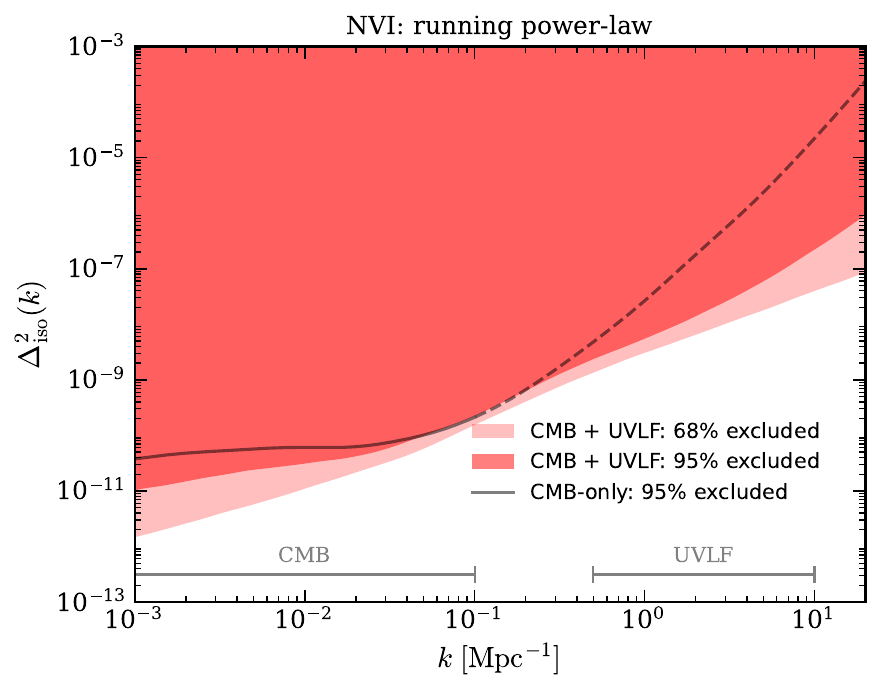}
\caption{$68\%$ and $95\%$ credible upper envelopes for the primordial neutrino velocity isocurvature power spectrum. The color shading follows the same convention as previously shown in Fig.~\ref{fig:isocurvature_envelopes}.}
    \label{fig:NVI_envelope}
\end{figure}

In Fig.~\ref{fig:NVI_envelope}, we present the upper bound for the NVI mode. Here, we restrict to the running parameterization for its simplicity, which
provides an adequate and representative description of the resulting constraints. On scales $k<0.01\,\mathrm{Mpc}^{-1}$, the CMB-only constraints are relatively weaker than those obtained from combined CMB+UVLF analysis. As discussed in the preceding paragraph, we associate this mild discrepancy with a steeper tilt or positive running which forces a lower amplitude on larger scales.

We note that, unlike the recent analysis of Ref.~\cite{Buckley:2025zgh}, we do not find a sharp improvement in the isocurvature constraint near $k\sim 1\,{\rm Mpc}^{-1}$ in Fig.~\ref{fig:isocurvature_envelopes}. In our UVLF-based analysis the bounds vary smoothly across this range. This behavior would be physically expected. By contrast, the compressed Lyman-$\alpha$ likelihood used in Ref.~\cite{Buckley:2025zgh} reduces the information in the 1D flux-power spectrum to constraints on the local amplitude and logarithmic slope of the linear matter power spectrum at a single pivot scale and redshift, approximately $k_p\sim 1\,{\rm Mpc}^{-1}$ and $z_p=3$. As also emphasized by the authors in Ref.~\cite{Buckley:2025zgh}, this approximation becomes less robust when the matter power spectrum changes slope rapidly near the pivot, and they explicitly associate the sharp transition in their bound around this scale with that limitation.

There is also a key methodological distinction. The compressed Lyman-$\alpha$ likelihood is calibrated on hydrodynamical simulations within standard adiabatic $\Lambda$CDM cosmologies and assumes that deviations can be captured by local changes in the amplitude and slope of $P_{\rm lin}(k,z)$ near the pivot. A fully dedicated Lyman-$\alpha$ analysis of isocurvature models would instead require targeted simulations or emulators. Notably, the EFT-based Lyman-$\alpha$ results compiled in Fig.~4 of Ref.~\cite{Gorghetto:2025uls} (from Ref.~\cite{Ivanov:2025pbu}) for $k^3$ isocurvature spectra show no corresponding dip over $k\simeq 0.04\text{--}10~{\rm Mpc}^{-1}$. Accordingly, our results are broadly consistent with existing Lyman-$\alpha$ constraints over the relevant range of scales.

\begin{figure}[t]
    \centering
\includegraphics[width=0.4\textwidth]{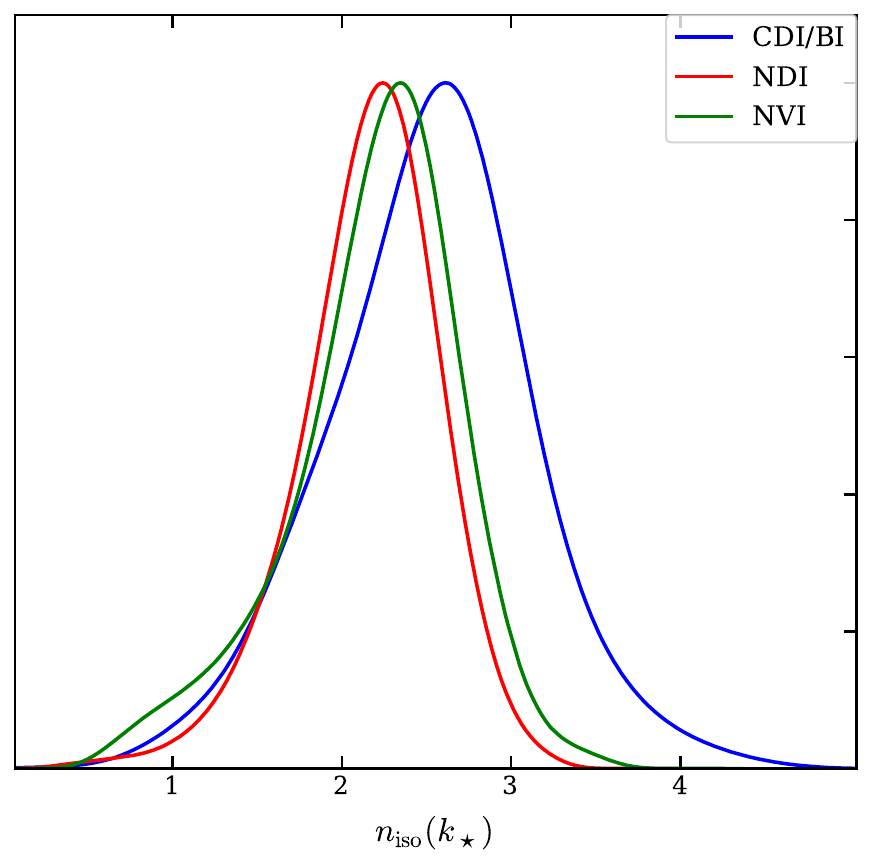}
\caption{Posterior distributions of the effective spectral index at a characteristic scale $k_\star\simeq0.4~\rm{Mpc}^{-1}$ for the CDI/BI, NDI, and NVI modes obtained using combined CMB and UVLF likelihoods.}
    \label{fig:alpha-niso2}
\end{figure}

From the figures, we note that the upper bounds on the isocurvature modes rise approximately as $k^n$ with $n\sim 2$ at high wavenumber. This sets the maximal rate at which the allowed isocurvature power can grow into the UVLF-sensitive regime ($0.5\rm{/Mpc}$$~<k<~$$ 10\rm{/Mpc}$)  while remaining consistent with the combined data. This behavior reflects the scale-dependent sensitivity of the datasets, with constraints on isocurvature power weakening toward smaller scales due to nonlinear mapping, astrophysical degeneracies, and transfer-function effects.

To characterize the representative growth of isocurvature power on small-scales allowed by the data for each isocurvature mode, we evaluate the effective slope or spectral index  using Eq.~\eqref{eq:scale-dependent-niso} at a characteristic scale $k_\star$. We choose $k_\star\simeq0.4~\rm{Mpc}^{-1}$, which roughly corresponds to the geometric mean of the two anchor scales used in our analysis, and is approximately the lower boundary of the UVLF-sensitive range. Thus, the effective spectral index is evaluated where the UVLF data begin to probe the isocurvature spectrum beyond the CMB-dominated scales. Fig.~\ref{fig:alpha-niso2} shows the posterior distribution of the effective spectral index $n_{\rm iso}(k_\star)$ across the CDI/BI, NDI, and NVI modes. Since the parameters $n_{\rm iso}$ and $\alpha_{\rm iso}$ exhibit a negative correlation based on Eq.~(\ref{eq:scale-dependent-niso}), data primarily constrain a correlated combination of these two parameters rather than each independently, in order to satisfy the sampled isocurvature amplitude at the higher anchor scale $k_2=3~\rm{Mpc}^{-1}$. Consequently, the derived posterior for the effective tilt $n_{\rm iso}(k_\star)$ provides a more physically informative characterization of the allowed scale dependence of the isocurvature spectrum.

We also find that CDI and BI permit slightly higher spectral tilts than NVI and NDI. This can be understood from the different matter transfer functions associated with each isocurvature mode~\cite{Bucher:1999re,Chluba_2013}. Matter isocurvature perturbations reside initially in a subdominant non-relativistic component during radiation domination, so their gravitational impact on the total density perturbation is suppressed until near matter--radiation equality. By contrast, neutrino density and velocity isocurvature perturbations initially reside in the relativistic sector and source metric perturbations more efficiently before equality. Despite these differences, the degeneracy direction remains nearly identical across all modes, suggesting that it is primarily determined by the scale sensitivity of the UVLF measurements rather than the detailed isocurvature dynamics.

\begin{figure}[h]
    \centering
    \includegraphics[width=0.48\textwidth]{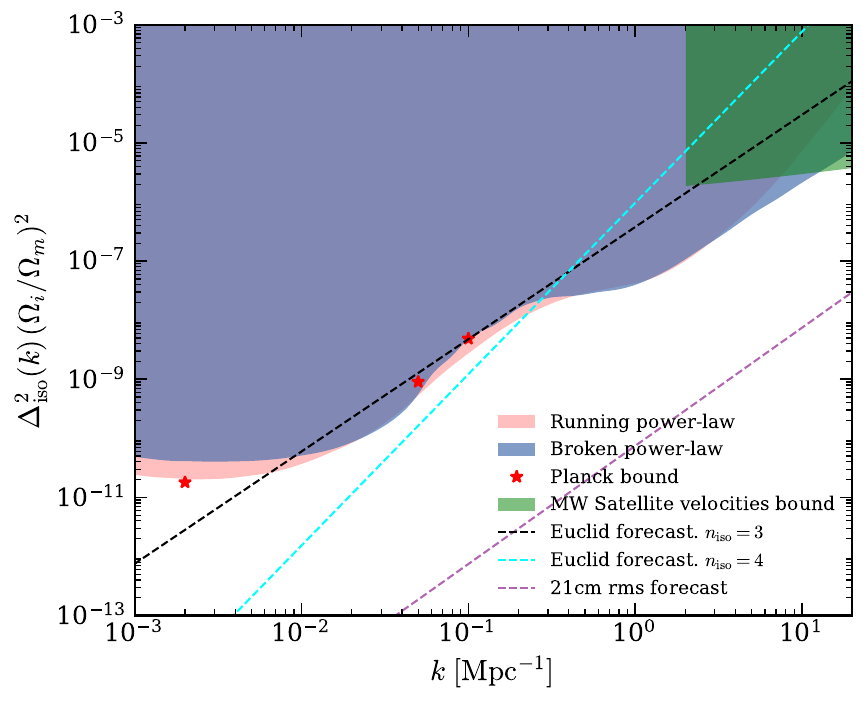}
    \caption{$95\%$ C.L. upper `excluded' limits on the primordial CDI/BI isocurvature power spectrum from this work, compared with selected current bounds and future forecasts. The blue and red shaded regions show the results obtained using the broken power-law and running parameterizations, respectively. Also shown are the current Planck and Milky Way satellite velocity bounds, together with representative Euclid and 21-cm forecasts. The vertical shaded bands mark the approximate scale ranges probed by the CMB and UVLF datasets. }
    \label{fig:CDI_95_compare}
\end{figure}

Finally, we conclude this section by presenting the $95\%$ C.L. constraints on
the CDI/BI isocurvature mode in Fig.~\ref{fig:CDI_95_compare} derived from both the broken power-law and running parameterizations, and comparing them with existing bounds and future forecasts. We restrict this comparison to the CDI/BI mode because most
published constraints and forecasts in the literature are quoted for matter
isocurvature, typically in the CDI normalization, with BI related by the usual
density-weighted rescaling. This makes CDI/BI the most direct and meaningful
case for comparison with previous work.

To give a sense of how the current constraints compare with other existing and future constraints, we sample results from the literature:
\begin{itemize}
\item Ref.~\cite{Planck:2018jri} analyzed the Planck PR3 likelihood to
constrain the isocurvature amplitude for various CDI models. Specifically, for the
uncorrelated AD+CDI with free $n_{{\rm {iso}}}$ (axion-II) model, 
they give the upper bound on isocurvature amplitude at three different scales $k_{\rm p}=0.002,~0.5,~0.1$ Mpc$^{-1}$. These are depicted in Fig.~\ref{fig:CDI_95_compare} as red star markers.

\item In Ref.~\cite{Esteban:2023xpk}, constraints on the small-scale matter power spectrum are obtained from the internal kinematics of Milky Way satellite galaxies, in particular the correlation between their velocity dispersions and sizes. These observables probe halo concentrations and formation times, and are sensitive to comoving scales in the range $4\,\mathrm{Mpc}^{-1} \lesssim k \lesssim 37\,\mathrm{Mpc}^{-1}$. The analysis constrains deviations of the matter power spectrum from scale invariance to be within a factor of $\sim 2$--$2.5$ over this range. 

We translate these bounds into approximate constraints on the isocurvature amplitude using the appropriate transfer functions, and represent the excluded amplitude with green shaded region. Since these constraints probe only relatively small scales, isocurvature spectra with a cutoff below $k \sim 5\,\mathrm{Mpc}^{-1}$ can evade them, whereas the UVLF constraints shown in Fig.~\ref{fig:CDI_95_compare} remain effective. As seen in the figure, the UVLF bounds are significantly stronger on overlapping scales, particularly for $k \lesssim 10\,\mathrm{Mpc}^{-1}$.

\item Ref.~\cite{Chung:2023syw} provided forecasts on CDI isocurvature power using Euclid's expected galaxy clustering measurements. We adopt their results for spectral indices $n_{\rm iso}=3$ and $n_{\rm iso}=4$, and display representative curves in our figure, shown as black and cyan dashed lines respectively. These curves represent the expected sensitivity of the Euclid towards detection of the isocurvature mode. These concur with those derived in Ref.~\cite{Euclid:2025hlc}. As shown in Ref.~\cite{Chung:2023syw}, the sensitivity for the Spec-S5 `MegaMapper' \cite{Schlegel:2019eqc} is expected to be an order of magnitude better than Euclid.

\item Ref.~\citep{Sekiguchi:2013lma} presented a forecast on CDI/BI from SKA based on the 21-cm power spectrum sourced by minihalos in the redshift range $z\in[5,20]$. Because the signal is dominated by small virialized objects entering through the halo mass function, these constraints probe highly nonlinear scales at $k\sim \mathcal{O}(100)\,\mathrm{Mpc}^{-1}$. The results of Ref.~\citep{Takeuchi:2013hza} are similar. In the plot, we denote this as the 21-cm rms constraint, shown as a dashed purple line constructed assuming an unbroken power-law with fiducial spectral index $n_{\rm iso}=3$. 
\end{itemize}

As seen in Fig.~\ref{fig:CDI_95_compare}, current UVLF-based bounds on scales $k<10\,\mathrm{Mpc}^{-1}$ are relatively stronger than those derived from Milky Way satellite galaxies. Forecasts from near-future surveys such as Euclid are also expected to be comparatively weaker on scales $k\gtrsim 1\,\mathrm{Mpc}^{-1}$. In contrast, probes of strongly nonlinear structure, such as SKA observations of 21-cm emission from minihalos, have the potential to improve constraints on the isocurvature amplitude by at least two orders of magnitude for large spectral indices. However, these projected gains are subject to significant uncertainties, as extensive marginalization over astrophysical nuisance parameters, baryonic models and limitations in modeling nonlinear structure formation may substantially weaken the resulting bounds.

More recently, Ref.~\cite{Garcia-Gallego:2026phh} constrained post-inflationary axion-like-particle (ALP) isocurvature amplitude $f_{\rm iso}$ at pivot scale $k_p$ using high-resolution Lyman-$\alpha$ forest data, adopting a fixed spectral index $n_{\rm iso}=4$. Since this high-blue-tilted spectrum scales as $\Delta_{\rm iso}^2\propto k^3$, the constraint is primarily driven by the smallest scales probed by Lyman-$\alpha$, corresponding roughly to $k\sim5$--$20\,{\rm Mpc}^{-1}$. Their bounds are comparable to those obtained in this work from UVLF on overlapping scales and, as noted by the authors, are slightly weaker compared to a separate UVLF-based analysis of the ALP in Ref.~\cite{Urrutia:2025fvp}. This highlights the strength of Lyman-$\alpha$ forest measurements as a probe of small-scale cosmological initial conditions, while our analysis extends to more general isocurvature spectra and initial conditions.

We do not include spectral distortion constraints or forecasts from Ref.~\citep{Chluba_2013}, as these are significantly weaker than the UVLF-based bounds over the range of scales $k \lesssim 10\,\mathrm{Mpc}^{-1}$. On far smaller scales, such constraints are likely to be superseded by those derived from dwarf satellites and dynamical reheating-based constraints \cite{Graham:2024hah}.

\section{Discussion}\label{Sec:discussion}

In this work, we have derived new constraints on primordial isocurvature perturbations by combining large-scale cosmological probes with small-scale structure information from the ultraviolet luminosity function of high-redshift galaxies. By jointly analyzing CMB, BAO, and Type Ia supernova data with recent \textit{HST} and \textit{JWST} UVLF measurements, we extended sensitivity to comoving wavenumbers up to $k \sim 10~\mathrm{Mpc}^{-1}$.

We have considered two complementary parameterizations, a broken power law and a power law with running, without fixing the spectral index a priori. We find that both parameterizations yield nearly identical $2\sigma$ credible envelopes for the isocurvature power spectrum over a broad range of scales from $0.001 \lesssim k \lesssim 5\,{\rm Mpc}^{-1}$. The inferred constraints in this range appear largely insensitive to assumptions about the deviations from power-law spectral shape, and can therefore be interpreted as robust, data-driven bounds on the scale dependence of primordial isocurvature.

Our results show that UVLF measurements provide a powerful probe of isocurvature perturbations. In particular, they significantly strengthen constraints on scales $k \sim 0.5$--$10~\mathrm{Mpc}^{-1}$.

Looking ahead, future improvements in both observations and modeling are expected to further enhance the constraining power of this approach. On the observational side, deeper and more precise spectroscopically-confirmed UVLF measurements from ongoing and upcoming \textit{JWST} programs will extend the redshift and luminosity reach of current datasets. Including currently available photometric samples from redshifts $z\sim17\text{--}25$~\cite{Perez-Gonzalez_2025,Castellano:2025vkt} may already provide some improvement though these require additional nuisance parameters for fitting~\cite{Urrutia:2025fvp}. 
In parallel, continued progress in high-resolution Lyman-$\alpha$ forest measurements and their modeling is expected to further strengthen constraints on small-scale power. On the theoretical side, improved understanding of the galaxy--halo connection and nonlinear structure formation will help reduce astrophysical uncertainties. In combination with forthcoming large-scale surveys, 21-cm observations, and developments in perturbation theory (EFTofLSS) and numerical simulations, these advances have the potential to significantly tighten bounds on primordial initial conditions across a wide range of scales.

\section*{Acknowledgement}
We thank Subhajit Ghosh for useful comments on the draft and discussion on this subject. This work was supported by the Department of Energy under Grant No.~DE-SC0025611 at Indiana University. This research was supported in part by Lilly Endowment, Inc., through its support for the Indiana University Pervasive Technology Institute. 

\begin{appendix}

\section{UVLF Pipeline}\label{App:uvlf_detail}

In this appendix, we briefly summarize the modeling of the UVLF, including the connection between cosmology and galaxy observables, the halo mass function (HMF) prescription, astrophysical parameterization of star formation, and the treatment of observational data and uncertainties. For further details, we refer the reader to Ref.~\cite{Sabti:2021xvh}, whose analysis pipeline forms the basis of the present work. This framework enables a consistent mapping from the linear matter power spectrum to observable galaxy number densities at high redshift.

\paragraph{From the matter power spectrum to halo abundances}

For a given cosmological model, the linear matter power spectrum $P(k,z)$ is computed using a Boltzmann solver (e.g.\ \textsf{CLASS}). The variance of the density field smoothed on mass scale $M$ is defined as
\begin{equation}
\sigma^2(M,z) = \int \frac{dk}{2\pi^2} \, k^2 P(k,z)\, W^2(kR),
\end{equation}
where $W(kR)$ is the window function and $R$ is the smoothing scale related to halo mass via
\begin{equation}
M = \frac{4\pi}{3}\,\bar{\rho}\,R^3,
\end{equation}
with $\bar{\rho}$ the mean matter density. In this work, we utilize a tophat window function.

The halo mass function is expressed in the excursion-set framework as
\begin{equation}
\frac{d n_h}{d\ln M_h} = -\frac{\bar{\rho}}{M_h}\, f(\nu)\, \frac{d\ln \sigma}{d\ln M_h},
\label{eq:hmf_general}
\end{equation}
where $n_h$ is halo number density, $M_h$ is the halo mass and $\nu \equiv \delta_c / \sigma(M_h,z)$ is the peak height and $\delta_c \simeq 1.686$ is the critical overdensity for collapse.

We adopt the Sheth--Tormen parameterization for the multiplicity function,
\begin{equation}
f(\nu) = A \sqrt{\frac{2q\nu^2}{\pi}} \left[1 + (q\nu^2)^{-p}\right] e^{-q\nu^2/2},
\end{equation}
with fiducial parameters $(A,p,q) = (0.3222,\,0.3,\,0.707)$. These values reproduce $\Lambda$CDM $N$-body results to $\mathcal{O}(20\%)$ accuracy over a broad range of masses and redshifts.

\paragraph{Halo--galaxy connection and star formation modeling}

Following the treatment in Ref.~\cite{Sabti:2021xvh}, we relate dark matter halos to galaxy ultraviolet luminosities through a phenomenological star-formation model. The star-formation rate (SFR) $\dot{M}_\star$ is assumed to follow the halo accretion rate $\dot{M}_h$,
\begin{equation}
\dot{M}_\star = f_\star(M_h,z)\, \dot{M}_h,
\end{equation}
where $f_\star$ is the star-formation efficiency (SFE).

We adopt the standard double power-law parameterization for the SFE,
\begin{equation}
f_\star(M_h,z) =
\frac{\epsilon_\star(z)}
{\left(M_h/M_c(z)\right)^{\alpha_\star(z)} + \left(M_h/M_c(z)\right)^{\beta_\star(z)}},
\label{eq:sfe_model}
\end{equation}
where:
\begin{itemize}
\item $\epsilon_\star$ sets the overall normalization,
\item $M_c$ is a characteristic halo mass scale,
\item $\alpha_\star < 0$ controls the low-mass slope,
\item $\beta_\star > 0$ controls the high-mass suppression.
\end{itemize}

This form captures the expected behavior that star formation is inefficient in both low-mass halos (due to photoheating from the UV background and supernova feedback) and high-mass halos (due to cooling inefficiencies in hot gas and AGN quenching).

The halo accretion rate is modeled as
\begin{equation}
\dot{M}_h = c\, M_h\, H(z),
\end{equation}
where $c$ is absorbed into $\epsilon_\star$.

The instantaneous UV luminosity is related to the SFR through
\begin{equation}
\dot{M}_\star = \kappa_{\rm UV}\, L_{\rm UV},
\end{equation}
with
\(
\kappa_{\rm UV} = 1.15 \times 10^{-28}\; M_\odot~\rm{s}~\mathrm{yr}^{-1}~\mathrm{erg}^{-1}
\)~\cite{Kennicutt:1998zb,Madau:1996hu,Madau:1997pg}. The conversion between luminosity and absolute magnitude is given by
\begin{equation}
\log_{10}\left(\frac{L_{\rm UV}}{\mathrm{erg\,s^{-1}}}\right)
= 0.4\,(51.63 - M_{\rm UV}).
\end{equation}

\paragraph{Construction of the UV luminosity function}

Finally, we connect the intrinsic UV luminosity of the galaxies with the HMF by defining UVLF $\Phi(M_\mathrm{UV},z)$ as 
\begin{equation}
\Phi(M_{\rm UV},z) = \frac{d n_g}{d M_{\rm UV}}
=
\frac{d n_h}{d M_h}\,
\frac{d M_h}{d M_{\rm UV}},
\end{equation}
where we assume one central galaxy per halo ($n_g=n_h$).

The stochasticity in the halo--luminosity relation $d M_h/d M_{\rm UV}$ is incorporated by convolving the UVLF with a Gaussian scatter in magnitude space,
\begin{equation}
\Phi_{\rm obs}(M_{\rm UV}) =
\int d M'_{\rm UV} \, \Phi(M'_{\rm UV})\,
\frac{\exp\left[-\frac{(M_{\rm UV}-M'_{\rm UV})^2}{2\sigma_{M_{\rm UV}}^2}\right]}{\sqrt{2\pi}\sigma_{M_{\rm UV}}}
\end{equation}
where we integrate over theoretical $M'_{\rm UV}=M'_{\rm UV}(M_h)$. 
This accounts for intrinsic variability and observational uncertainties, and leads to Eddington bias at the bright end~\cite{eddington_formula_1913}.

To compare with binned UVLF data, we consider the integral in the above expression over the UV-magnitude bin-width $\Delta M_{\rm UV}$ and entire possible range of halo mass $M_h$.

\paragraph{Astrophysical nuisance parameters}

The UVLF modeling introduces the following nuisance parameters:
\begin{equation}
\{\alpha_\star,\ \beta_\star,\ \epsilon_\star,\ M_c,\ \sigma_{M_{\rm UV}}\}.
\end{equation}

These parameters exhibit partial degeneracies. For instance, $\epsilon_\star$ and the overall HMF normalization both affect the amplitude of the UVLF, while $\sigma_{M_{\rm UV}}$ primarily impacts the bright end. Wide magnitude coverage is therefore essential to break these degeneracies.

\paragraph{UVLF data and likelihood construction}

We use UVLF measurements from deep-field surveys, including:
\begin{itemize}
\item \textit{Hubble Space Telescope (\textit{HST})} data at $z \sim 4$--$8$,
\item \textit{James Webb Space Telescope (\textit{JWST})} data at $z \sim 9$--$13$.
\end{itemize}

Each data point consists of $(z, M_{\rm UV}, \Phi)$ with associated uncertainties.

To account for cosmic variance, we adopt a conservative error prescription,
\begin{equation}
\sigma_i^2 = \max\left(\sigma_{{\rm stat},i}^2,\ (f_{\rm CV}\,\Phi_i)^2\right),
\end{equation}
with $f_{\rm CV} \simeq 0.2$.

The likelihood is constructed assuming Gaussian errors,
\begin{equation}
-2\ln \mathcal{L} =
\sum_i
\left(
\frac{\Phi_{{\rm model},i} - \Phi_{{\rm data},i}}
{\sigma_i}
\right)^2.
\end{equation}

We additionally account for the following effects discussed in Ref.~\cite{Sabti:2021xvh}
\begin{itemize}
\item dust attenuation corrections (e.g.\ IRX--$\beta$ relations),
\item Alcock--Paczy\'nski rescaling when comparing to data assuming a fiducial cosmology.
\end{itemize}

\paragraph{Priors on UVLF nuisance parameters}

The UVLF modeling introduces a set of astrophysical nuisance parameters governing the halo--galaxy connection and intrinsic scatter. We adopt broad, uniform priors on these parameters, chosen to encompass the range of values typically inferred in $\Lambda$CDM-based analyses of high-redshift galaxy formation.

The full set of UVLF nuisance parameters and their prior ranges are summarized in Tab.~\ref{tab:uvlf_priors}. In addition to the standard parameters, we include an extra degree of freedom $\alpha_{\star,\rm s}$ to allow for mild deviations in the low-mass scaling of the star-formation efficiency that provides a better fit to the \textit{JWST} datasets within the $\Lambda$CDM framework.
\begin{table}[h]
\centering
\renewcommand{\arraystretch}{1.3}
\begin{tabular}{|c|c|}
\hline
Parameter & Prior \\
\hline
$\alpha_\star$ & $\mathcal{U}[-2.0,\,-0.01]$ \\
$\beta_\star$ & $\mathcal{U}[0.01,\,3]$ \\
$\log_{10}\epsilon_\star$ & $\mathcal{U}[-3,\,3]$ \\
$\log_{10}(M_c/M_\odot)$ & $\mathcal{U}[7,\,15]$ \\
$\sigma_{M_{\rm UV}}$ & $\mathcal{U}[0.001,\,3]$ \\
$\alpha_{\star,s}$ & $\mathcal{U}[-0.5,\,0.5]$ \\
\hline
\end{tabular}
\caption{Flat priors on the UVLF astrophysical nuisance parameters used in the analysis.}
\label{tab:uvlf_priors}
\end{table}

The prior ranges are selected to allow sufficient flexibility to capture uncertainties in star formation and feedback processes, while avoiding unphysical regions of parameter space (e.g.\ excessively steep slopes or unrealistically large scatter). In practice, certain degeneracies among nuisance parameters (e.g.\ between $\epsilon_\star$ and the overall normalization of the halo mass function) are partially broken by the wide magnitude coverage of the UVLF data.

\section{Isocurvature Spectrum Reconstruction}
\label{app:spectrum_reconstruction}

In this appendix we describe in detail the procedure used to reconstruct the
isocurvature spectral envelopes shown in Sec.~\ref{Sec:Results}. The purpose of this reconstruction is to provide a compact visual summary of the family of isocurvature spectra allowed by the data within the assumed spectral parameterizations. We are not reconstructing the power spectrum by freely fitting independent values at each scale. Instead, we impose a parametric form that restricts how the spectrum can vary with $k$.

\subsection{Broken Power-Law Parameterization}

For the broken power-law analysis, the break scale $k_c$ is treated as a fixed hyperparameter during each MCMC run. Thus, for each fixed value of $k_c$, the
MCMC samples are drawn from the conditional posterior
\begin{equation}
p(\Theta \mid D,k_c),
\end{equation}
where $D$ denotes the combined dataset/likelihood and $\Theta$ denotes the full set of
sampled parameters, including the standard cosmological parameters, nuisance
parameters, and the two sampled isocurvature amplitudes
\begin{equation}
\mathcal{A}_{1}\equiv \mathcal{A}_{\rm iso}(k_1),
\qquad
\mathcal{A}_{2}\equiv \mathcal{A}_{\rm iso}(k_2).
\end{equation}

For a given value of $k_c$, the sampled pair
$\{\mathcal{A}_{1},\mathcal{A}_{2}\}$ uniquely determines the spectral shape
within the assumed broken power-law template. In the implementation used in this
work, the spectrum is reconstructed as a power law below the break and as a
constant plateau above the break. When both anchor scales lie below the plateau, the effective tilt is determined
by
\begin{equation}
n_{{\rm iso},s}-1
=
\frac{
\ln\left(\mathcal{A}_{2,s}/\mathcal{A}_{1,s}\right)
}{
\ln\left(k_2/k_1\right)
},
\label{eq:derived_tilt}
\end{equation}
and the amplitude at the break is obtained from
\begin{equation}
\mathcal{A}_{c,s}
=
\mathcal{A}_{1,s}
\left(\frac{k_c}{k_1}\right)^{n_{{\rm iso},s}-1}
=
\mathcal{A}_{2,s}
\left(\frac{k_c}{k_2}\right)^{n_{{\rm iso},s}-1}
\label{eq:derived_Akc}
\end{equation}
where the subscript $s$ denotes a sampled instance.

More generally, if $k_2$  lies on the plateau side of the
break,
\begin{equation}
\mathcal{A}_{c,s}=\mathcal{A}_{2,s},\qquad n_{\rm iso,s}=\frac{
\ln\left(\mathcal{A}_{2,s}/\mathcal{A}_{1,s}\right)
}{
\ln\left(k_c/k_1\right)
}.
\end{equation}
Thus the sampled amplitudes at the anchor
scales determine the unique member of the broken power-law family associated
with that posterior sample and that fixed value of $k_c$.

In the analysis we repeat this procedure for a discrete set of fixed break
scales in the range $k_c\in [100,~0.001]~{\rm Mpc}^{-1}$.
These values span the range from large scales relevant to the CMB to small
scales relevant for the UV luminosity function and beyond. 

\paragraph{Combination of fixed-$k_c$ chains}
Because each value of $k_c$ defines a separate MCMC analysis, the
posterior samples from each MCMC analysis are independent samples from different
conditional distributions,
\begin{equation}
p(\Theta \mid D,k_{c,j}),
\end{equation}
where $j$ labels the discrete break scale. Therefore, combining the chains
requires specifying the statistical meaning of the combined ensemble.

Since the data do not show any preference for a break-scale, in the reconstruction shown in the main text, we form an equal-weight mixture
over the discrete set of break scales. That is, each
fixed-$k_c$ analysis contributes the same number of posterior samples to the
combined ensemble. The resulting distribution over reconstructed spectra may be written schematically as
\begin{equation}
p_{\rm mix}\!\left[\Delta^2_{\rm iso}(k)\mid D\right]
=
\sum_{j=1}^{N_c}
w_j\,
p\!\left[\Delta^2_{\rm iso}(k)\mid D,k_{c,j}\right],
\label{eq:mixture_distribution}
\end{equation}
with
\(w_j = 1/N_c\)
where $N_c$ is the total number of fixed $k_c$ values chosen for the analyses. 
Here
$p[\Delta^2_{\rm iso}(k)\mid D,k_{c,j}]$ denotes the distribution of
reconstructed spectra generated by the posterior samples from the MCMC chain with a fixed break scale $k_{c,j}$.

This equal-weight mixture should be interpreted as a summary over the chosen
discrete family of broken power-law templates. It is not an evidence-weighted
Bayesian model average over $k_c$. A fully Bayesian marginalization over the
break scale would require assigning a prior $p(k_c)$ and weighting each
fixed-$k_c$ model by its posterior model probability. Our scheme is thus equivalent (and applicable) to the scenario where the posterior is flat over the entire range of sampled $k_c$ values.

\paragraph{Pointwise posterior quantiles}

For each fixed value $k_{c,j}$, we draw a representative (randomly sorted) subset of converged
posterior samples from the corresponding MCMC chain. For each posterior sample
$s$, we reconstruct the continuous spectrum
\begin{equation}
\Delta^2_{{\rm iso},s}(k;k_{c,j})
\end{equation}
using Eq.~\eqref{eq:broken-power-law} on a common logarithmically spaced fine-grid,
\(
\mathcal{K} = \{k_\ell\}_{\ell=1}^{N_k}.
\)
At each grid point $k_\ell$, the combined ensemble consists of the values
\begin{equation}
\left\{
\Delta^2_{{\rm iso},s}(k_\ell;k_{c,j})
\right\}_{s,j}.
\end{equation}
The pointwise $q$-quantile envelope is then defined by the empirical quantile
\begin{equation}
Q_q(k_\ell)
=
{\rm Quantile}_{q}
\left[
\left\{
\Delta^2_{{\rm iso},s}(k_\ell;k_{c,j})
\right\}_{s,j}
\right].
\label{eq:pointwise_quantile}
\end{equation}
For example, the $95\%$ ($68\%$) upper envelope is
\begin{equation}
\Delta^2_{{\rm iso},95 (68)}(k_\ell)
=
Q_{0.95 (68)}(k_\ell).
\end{equation}

Here, the term ``pointwise'' is important. Equation~\eqref{eq:pointwise_quantile}
defines the credible upper limit separately at each value of $k_\ell$. Therefore
the envelope should not be interpreted as a simultaneous credible band for an
entire spectral function. In other words, the statement
\begin{equation}
{\rm Pr}
\left[
\Delta^2_{\rm iso}(k_\ell)
\leq
Q_{0.95}(k_\ell)
\mid D
\right]
=
0.95
\end{equation}
holds separately for each grid point $k_\ell$ within the mixture ensemble, but
this does not imply that $95\%$ of full spectral curves lie below
$Q_{0.95}(k)$ for all $k$ simultaneously.

For visualization only, the quantile curves/envelope are mildly smoothed in
$\log \Delta^2_{\rm iso}$ as a function of $\log k$. This smoothing is applied
after the empirical quantiles have been computed and is used only to suppress
finite-sampling noise in the plotted envelope.

The reconstructed broken power-law envelope should be interpreted as a compact
summary of the allowed spectra within the specific model class defined by
Eq.~\eqref{eq:broken-power-law}. It is not a reconstruction of a `free' primordial power spectrum with independently sampled amplitudes in narrow $k$-bins. 

This distinction is particularly relevant when interpreting the upper envelope.
Since different values of $k_c$ contribute to the mixture, the pointwise upper
bound at different wavenumbers (or comoving scales) may be saturated by spectra originating from
different fixed-$k_c$ analyses. Therefore, the envelope traces the allowed
range of power scale by scale across the family of templates, rather than
representing a single physical spectrum that is jointly preferred at all
wavenumbers. In this sense, the broken power-law envelope is best viewed as a
near-maximal pointwise summary of the allowed isocurvature power within the
chosen template family.

\subsection{Running Parameterization}

For comparison, the running isocurvature parameterization is reconstructed in
a similar manner to the broken power-law case, but without introducing a break
scale. The spectrum is specified by an amplitude, a spectral index, and a
running. In our sampling strategy, we sample the amplitudes at two anchor scales
$k_{1,2}$ together with the spectral index, and derive the running using the
assumed functional form in Eq.~\eqref{eq:powerlaw_running}.

We perform two separate MCMC runs for the running
model, differing only in the choice of the lower anchor scale $k_1$. Specifically,
we use $k_1=0.01$~Mpc$^{-1}$ and $k_1=0.05$~Mpc$^{-1}$, while keeping $k_2=3$~Mpc$^{-1}$ fixed. Each run therefore
samples from a conditional posterior of the form
\begin{equation}
p(\Theta \mid D, k_{1,i}),
\end{equation}
where $i$ labels the choice of anchor scale.

Each posterior sample from a given run defines a continuous reconstructed spectrum,
\begin{equation}
\Delta^2_{{\rm iso},s}(k;k_{1,i}),
\end{equation}
which is evaluated on the same logarithmic $k$ grid $\mathcal{K}$.

Since the two analyses correspond to different choices of anchor scale, their
posterior samples are drawn from different conditional distributions. To construct
a combined summary, we form an equal-weight mixture of the reconstructed spectra
from the two runs, analogous to Eq.~\eqref{eq:mixture_distribution}.

At each grid point $k_\ell$, the combined ensemble consists of
\(
\left[
\Delta^2_{{\rm iso},s}(k_\ell;k_{1,i})
\right]_{s,i}.
\)
The pointwise $q$-quantile envelope is then defined as
\begin{equation}
Q^{\rm run}_q(k_\ell) = {\rm Quantile}_{q}
\left[
\left\{ \Delta^2_{{\rm iso},s}(k_\ell;k_{1,i})\right\}_{s,i}
\right].
\end{equation}

As in the broken power-law case, these are pointwise credible intervals and do not represent simultaneous bounds on the full spectral function.

\end{appendix}

\bibliographystyle{JHEP2.bst}
\bibliography{newref,refs}

@article{Sabti:2021xvh,
    author = "Sabti, Nashwan and Mu{\~n}oz, Julian B. and Blas, Diego",
    title = "{Galaxy luminosity function pipeline for cosmology and astrophysics}",
    eprint = "2110.13168",
    archivePrefix = "arXiv",
    primaryClass = "astro-ph.CO",
    reportNumber = "KCL-2021-74",
    doi = "10.1103/PhysRevD.105.043518",
    journal = "Phys. Rev. D",
    volume = "105",
    number = "4",
    pages = "043518",
    year = "2022"
}

@ARTICLE{donnan2024jwst,
       author = {{Donnan}, C.~T. and {McLure}, R.~J. and {Dunlop}, J.~S. and {McLeod}, D.~J. and {Magee}, D. and {Arellano-Cordova}, K.~Z. and {Barrufet}, L. and {Begley}, R. and {Bowler}, R.~A.~A. and {Carnall}, A.~C. and {Cullen}, F. and {Ellis}, R.~S. and {Fontana}, A. and {Illingworth}, G.~D. and {Grogin}, N.~A. and {Hamadouche}, M.~L. and {Koekemoer}, A.~M. and {Liu}, F.-Y. and {Mason}, C. and {Santini}, P. and {Stanton}, T.~M.},
        title = "{JWST PRIMER: a new multifield determination of the evolving galaxy UV luminosity function at redshifts z $\sim$ 9 -- 15}",
      journal = {mnras},
         year = 2024,
        month = sep,
       volume = {533},
       number = {3},
        pages = {3222-3237},
          doi = {10.1093/mnras/stae2037},
archivePrefix = {arXiv},
       eprint = {2403.03171},
 primaryClass = {astro-ph.GA},
}

@article{Oesch_2018,
doi = {10.3847/1538-4357/aab03f},
url = {https://doi.org/10.3847/1538-4357/aab03f},
year = {2018},
month = {mar},
publisher = {The American Astronomical Society},
volume = {855},
number = {2},
pages = {105},
author = {Oesch, P. A. and Bouwens, R. J. and Illingworth, G. D. and Labbé, I. and Stefanon, M.},
title = {The Dearth of z $\sim$ 10 Galaxies in All HST Legacy Fields-The Rapid Evolution of the Galaxy Population in the First 500 Myr*},
journal = {The Astrophysical Journal},
}

@article{Bouwens_2021,
doi = {10.3847/1538-3881/abf83e},
url = {https://doi.org/10.3847/1538-3881/abf83e},
year = {2021},
month = {jul},
publisher = {The American Astronomical Society},
volume = {162},
number = {2},
pages = {47},
author = {Bouwens, R. J. and Oesch, P. A. and Stefanon, M. and Illingworth, G. and Labbé, I. and Reddy, N. and Atek, H. and Montes, M. and Naidu, R. and Nanayakkara, T. and Nelson, E. and Wilkins, S.},
title = {New Determinations of the UV Luminosity Functions from z $\sim$ 9 to 2 Show a Remarkable Consistency with Halo Growth and a Constant Star Formation Efficiency},
journal = {The Astronomical Journal},
}

@article{Madau:1997pg,
    author = "Madau, Piero and Pozzetti, Lucia and Dickinson, Mark",
    title = "{The Star formation history of field galaxies}",
    eprint = "astro-ph/9708220",
    archivePrefix = "arXiv",
    doi = "10.1086/305523",
    journal = "Astrophys. J.",
    volume = "498",
    pages = "106",
    year = "1998"
}

@article{Kennicutt:1998zb,
    author = "Kennicutt, Jr., Robert C.",
    title = "{Star formation in galaxies along the Hubble sequence}",
    eprint = "astro-ph/9807187",
    archivePrefix = "arXiv",
    reportNumber = "STEWARD-1454, STEWARD-OBSERVATORY-PREPRINT-1454",
    doi = "10.1146/annurev.astro.36.1.189",
    journal = "Ann. Rev. Astron. Astrophys.",
    volume = "36",
    pages = "189--231",
    year = "1998"
}

@article{Madau:1996hu,
    author = "Madau, Piero",
    editor = "Holt, Stephen S. and Mundy, Lee G.",
    title = "{Cosmic star formation history}",
    eprint = "astro-ph/9612157",
    archivePrefix = "arXiv",
    doi = "10.1063/1.52821",
    journal = "AIP Conf. Proc.",
    volume = "393",
    number = "1",
    pages = "481",
    year = "1997"
}

@article{Pan-STARRS1:2017jku,
    author = "Scolnic, D. M. and others",
    collaboration = "Pan-STARRS1",
    title = "{The Complete Light-curve Sample of Spectroscopically Confirmed SNe Ia from Pan-STARRS1 and Cosmological Constraints from the Combined Pantheon Sample}",
    eprint = "1710.00845",
    archivePrefix = "arXiv",
    primaryClass = "astro-ph.CO",
    doi = "10.3847/1538-4357/aab9bb",
    journal = "Astrophys. J.",
    volume = "859",
    number = "2",
    pages = "101",
    year = "2018"
}

@article{Jones:2017udy,
    author = "Jones, D. O. and others",
    title = "{Measuring Dark Energy Properties with Photometrically Classified Pan-STARRS Supernovae. II. Cosmological Parameters}",
    eprint = "1710.00846",
    archivePrefix = "arXiv",
    primaryClass = "astro-ph.CO",
    doi = "10.3847/1538-4357/aab6b1",
    journal = "Astrophys. J.",
    volume = "857",
    number = "1",
    pages = "51",
    year = "2018"
}

@article{brinckmann_montepython_2019,
	title = {{MontePython} 3: {Boosted} {MCMC} sampler and other features},
	volume = {24},
	issn = {2212-6864},
	url = {https://www.sciencedirect.com/science/article/pii/S2212686418302309},
	doi = {https://doi.org/10.1016/j.dark.2018.100260},
	journal = {Physics of the Dark Universe},
	author = {Brinckmann, Thejs and Lesgourgues, Julien},
	year = {2019},
	pages = {100260},
}

@article{lewis_getdist_2025,
	title = {{GetDist}: a {Python} package for analysing {Monte} {Carlo} samples},
	volume = {2025},
	url = {https://doi.org/10.1088/1475-7516/2025/08/025},
	doi = {10.1088/1475-7516/2025/08/025},
	number = {08},
	journal = {Journal of Cosmology and Astroparticle Physics},
	author = {Lewis, Antony},
	month = aug,
	year = {2025},
	note = {Publisher: IOP Publishing},
	pages = {025},
}

@article{eddington_formula_1913,
	title = {On a {Formula} for {Correcting} {Statistics} for the {Effects} of a known {Probable} {Error} of {Observation}},
	volume = {73},
	issn = {0035-8711},
	url = {https://doi.org/10.1093/mnras/73.5.359},
	doi = {10.1093/mnras/73.5.359},
	number = {5},
	journal = {Monthly Notices of the Royal Astronomical Society},
	author = {Eddington, A. S.},
	month = mar,
	year = {1913},
	note = {\_eprint: https://academic.oup.com/mnras/article-pdf/73/5/359/2946057/mnras73-0359.pdf},
	pages = {359--360},
}

@article{Graham:2024hah,
    author = "Graham, Peter W. and Ramani, Harikrishnan",
    title = "{Constraints on dark matter from dynamical heating of stars in ultrafaint dwarfs. II. Substructure and the primordial power spectrum}",
    eprint = "2404.01378",
    archivePrefix = "arXiv",
    primaryClass = "hep-ph",
    doi = "10.1103/PhysRevD.110.075012",
    journal = "Phys. Rev. D",
    volume = "110",
    number = "7",
    pages = "075012",
    year = "2024"
}

@article{Buckley:2025zgh,
    author = "Buckley, Matthew R. and Du, Peizhi and Fernandez, Nicolas and Weikert, Mitchell J.",
    title = "{General constraints on isocurvature from the CMB and Ly-{\ensuremath{\alpha}} forest}",
    eprint = "2502.20434",
    archivePrefix = "arXiv",
    primaryClass = "astro-ph.CO",
    doi = "10.1088/1475-7516/2025/12/006",
    journal = "JCAP",
    volume = "12",
    pages = "006",
    year = "2025"
}

@article{Bae:2026hly,
    author = "Bae, Kyu Jung and Cheong, Dhong Yeon and Gong, Jinn-Ouk and Harigaya, Keisuke and Shin, Chang Sub",
    title = "{Kinetic Isocurvature Perturbation}",
    eprint = "2603.22394",
    archivePrefix = "arXiv",
    primaryClass = "hep-ph",
    reportNumber = "APCTP-Pre2026-006",
    month = "3",
    year = "2026"
}

@article{Chung:2017uzc,
    author = "Chung, Daniel J. H. and Upadhye, Amol",
    title = "{Search for strongly blue axion isocurvature}",
    eprint = "1711.06736",
    archivePrefix = "arXiv",
    primaryClass = "astro-ph.CO",
    doi = "10.1103/PhysRevD.98.023525",
    journal = "Phys. Rev. D",
    volume = "98",
    number = "2",
    pages = "023525",
    year = "2018"
}

@ARTICLE{Shadab_uvlf,
       author = {Kar, Abhijnan and Alam, Shadab and Silk, Joseph},
        title = "{Beyond Extreme Burstiness: Evolving Star Formation Efficiency as the Key to Early Galaxy Abundance}",
      journal = {\apj},
         year = 2026,
        month = jan,
       volume = {996},
       number = {1},
          eid = {103},
        pages = {103},
          doi = {10.3847/1538-4357/ae1f8b},
archivePrefix = {arXiv},
       eprint = {2507.20606},
 primaryClass = {astro-ph.GA},
       adsurl = {https://ui.adsabs.harvard.edu/abs/2026ApJ...996..103K}
}

@article{Co:2025hbi,
    author = "Co, Raymond T. and Lam, Siu Cheung and Tadepalli, Sai Chaitanya and Takahashi, Tomo",
    title = "{Reviving sub-keV warm dark matter: a UVLF-based analysis}",
    eprint = "2512.16987",
    archivePrefix = "arXiv",
    primaryClass = "astro-ph.CO",
    month = "12",
    year = "2025",
    journal="JCAP"
}

@article{Schlegel:2019eqc,
    author = "Schlegel, David J. and others",
    title = "{Astro2020 APC White Paper: The MegaMapper: a z {\ensuremath{>}} 2 Spectroscopic Instrument for the Study of Inflation and Dark Energy}",
    eprint = "1907.11171",
    archivePrefix = "arXiv",
    primaryClass = "astro-ph.IM",
    reportNumber = "FERMILAB-FN-1081-AE-SCD",
    journal = "Bull. Am. Astron. Soc.",
    volume = "51",
    number = "7",
    pages = "229",
    year = "2019"
}

@article{Sekiguchi:2013lma,
    author = "Sekiguchi, Toyokazu and Tashiro, Hiroyuki and Silk, Joseph and Sugiyama, Naoshi",
    title = "Cosmological signatures of tilted isocurvature perturbations: reionization and 21cm fluctuations",
    eprint = "1311.3294",
    archivePrefix = "arXiv",
    primaryClass = "astro-ph.CO",
    reportNumber = "HIP-2013-26-TH",
    doi = "10.1088/1475-7516/2014/03/001",
    journal = "JCAP",
    volume = "03",
    pages = "001",
    year = "2014"
}

@article{Takeuchi:2013hza,
    author = "Takeuchi, Yoshitaka and Chongchitnan, Sirichai",
    title = "{Constraining isocurvature perturbations with the 21 cm emission from minihaloes}",
    eprint = "1311.2585",
    archivePrefix = "arXiv",
    primaryClass = "astro-ph.CO",
    doi = "10.1093/mnras/stu059",
    journal = "Mon. Not. Roy. Astron. Soc.",
    volume = "439",
    number = "1",
    pages = "1125--1135",
    year = "2014"
}

@article{Chluba_2013,
	doi = {10.1093/mnras/stt1129},
 	url = {https://doi.org/10.1093%2Fmnras%2Fstt1129},
 	year = 2013,
	month = {jul},
 	publisher = {Oxford University Press (OUP)},
  	volume = {434},
  	number = {2},
  	pages = {1619--1635},
  	author = {J. Chluba and D. Grin},
  	title = {{CMB} spectral distortions from small-scale isocurvature fluctuations},
         eprint = "1304.4596",
  	journal = {Monthly Notices of the Royal Astronomical Society}
}

@article{Euclid:2025hlc,
    author = "Finelli, F. and others",
    collaboration = "Euclid",
    title = "{Euclid preparation: Expected constraints on initial conditions}",
    eprint = "2507.15819",
    archivePrefix = "arXiv",
    primaryClass = "astro-ph.CO",
    month = "7",
    year = "2025"
}

@article{AtacamaCosmologyTelescope:2025blo,
    author = "Louis, Thibaut and others",
    collaboration = "Atacama Cosmology Telescope",
    title = "{The Atacama Cosmology Telescope: DR6 power spectra, likelihoods and {\ensuremath{\Lambda}}CDM parameters}",
    eprint = "2503.14452",
    archivePrefix = "arXiv",
    primaryClass = "astro-ph.CO",
    reportNumber = "FERMILAB-PUB-25-0071-PPD",
    doi = "10.1088/1475-7516/2025/11/062",
    journal = "JCAP",
    volume = "11",
    pages = "062",
    year = "2025"
}

@article{Seckel:1985tj,
    author = "Seckel, D. and Turner, Michael S.",
    title = "{Isothermal Density Perturbations in an Axion Dominated Inflationary Universe}",
    reportNumber = "FERMILAB-PUB-85-087-A",
    doi = "10.1103/PhysRevD.32.3178",
    journal = "Phys. Rev. D",
    volume = "32",
    pages = "3178",
    year = "1985"
}

@article{Lyth:2001nq,
    author = "Lyth, David H. and Wands, David",
    title = "{Generating the curvature perturbation without an inflaton}",
    eprint = "hep-ph/0110002",
    archivePrefix = "arXiv",
    reportNumber = "PU-RCG-01-33",
    doi = "10.1016/S0370-2693(01)01366-1",
    journal = "Phys. Lett. B",
    volume = "524",
    pages = "5--14",
    year = "2002"
}

@article{Enqvist:2001zp,
    author = "Enqvist, Kari and Sloth, Martin S.",
    title = "{Adiabatic CMB perturbations in pre - big bang string cosmology}",
    eprint = "hep-ph/0109214",
    archivePrefix = "arXiv",
    reportNumber = "HIP-2001-51-TH",
    doi = "10.1016/S0550-3213(02)00043-3",
    journal = "Nucl. Phys. B",
    volume = "626",
    pages = "395--409",
    year = "2002"
}

@article{Moroi:2001ct,
    author = "Moroi, Takeo and Takahashi, Tomo",
    title = "{Effects of cosmological moduli fields on cosmic microwave background}",
    eprint = "hep-ph/0110096",
    archivePrefix = "arXiv",
    reportNumber = "TU-632",
    doi = "10.1016/S0370-2693(01)01295-3",
    journal = "Phys. Lett. B",
    volume = "522",
    pages = "215--221",
    year = "2001",
    note = "[Erratum: Phys.Lett.B 539, 303--303 (2002)]"
}

@article{Polarski:1994rz,
    author = "Polarski, David and Starobinsky, Alexei A.",
    title = "{Isocurvature perturbations in multiple inflationary models}",
    eprint = "astro-ph/9404061",
    archivePrefix = "arXiv",
    reportNumber = "YITP-U-94-7",
    doi = "10.1103/PhysRevD.50.6123",
    journal = "Phys. Rev. D",
    volume = "50",
    pages = "6123--6129",
    year = "1994"
}

@article{Byrnes:2006fr,
    author = "Byrnes, Christian T. and Wands, David",
    title = "{Curvature and isocurvature perturbations from two-field inflation in a slow-roll expansion}",
    eprint = "astro-ph/0605679",
    archivePrefix = "arXiv",
    doi = "10.1103/PhysRevD.74.043529",
    journal = "Phys. Rev. D",
    volume = "74",
    pages = "043529",
    year = "2006"
}

@article{Chung:1998zb,
    author = "Chung, Daniel J. H. and Kolb, Edward W. and Riotto, Antonio",
    title = "{Superheavy dark matter}",
    eprint = "hep-ph/9802238",
    archivePrefix = "arXiv",
    reportNumber = "FERMILAB-PUB-98-021-A, CERN-TH-98-37, OUTP-98-02-P, FERMILAB-FERMILAB-PUB-98-021-A, CERN-CERN-TH-98-37, OXFORD-U. --OUTP-98-02-P",
    doi = "10.1103/PhysRevD.59.023501",
    journal = "Phys. Rev. D",
    volume = "59",
    pages = "023501",
    year = "1998"
}

@article{Freese:2023fcr,
    author = "Freese, Katherine and Winkler, Martin Wolfgang",
    title = "{Dark matter and gravitational waves from a dark big bang}",
    eprint = "2302.11579",
    archivePrefix = "arXiv",
    primaryClass = "astro-ph.CO",
    reportNumber = "UTWI-06-2023, NORDITA-2023-003",
    doi = "10.1103/PhysRevD.107.083522",
    journal = "Phys. Rev. D",
    volume = "107",
    number = "8",
    pages = "083522",
    year = "2023"
}

@article{Tadepalli:2026mdc,
    author = "Tadepalli, Sai Chaitanya",
    title = "{Axions on a Hyperbolic Ride: Geometric Suppression of CMB Isocurvature and a Blue-Tilted Spectrum}",
    eprint = "2601.22221",
    archivePrefix = "arXiv",
    primaryClass = "hep-ph",
    month = "1",
    year = "2026"
}

@article{Perez-Gonzalez_2025,
doi = {10.3847/1538-4357/adf8c9},
url = {https://doi.org/10.3847/1538-4357/adf8c9},
year = {2025},
month = {sep},
publisher = {The American Astronomical Society},
volume = {991},
number = {2},
pages = {179},
author = {Pérez-González, Pablo G. and Östlin, Göran and Costantin, Luca and Melinder, Jens and Finkelstein, Steven L. and Somerville, Rachel S. and Annunziatella, Marianna and Álvarez-Márquez, Javier and Colina, Luis and Dekel, Avishai and Ferguson, Henry C. and Li, Zhaozhou and Yung, L. Y. Aaron and Bagley, Micaela B. and Boogaard, Leindert A. and Burgarella, Denis and Calabrò, Antonello and Caputi, Karina I. and Cheng, Yingjie and Dickinson, Mark and Eckart, Andreas and Giavalisco, Mauro and Gillman, Steven and Greve, Thomas R. and Hamed, Mahmoud and Hathi, Nimish P. and Hjorth, Jens and Huertas-Company, Marc and Kartaltepe, Jeyhan S. and Koekemoer, Anton M. and Kokorev, Vasily and Labiano, Álvaro and Langeroodi, Danial and Leung, Gene C. K. and Natarajan, Priyamvada and Papovich, Casey and Peissker, Florian and Pentericci, Laura and Pirzkal, Nor and Rinaldi, Pierluigi and van der Werf, Paul and Walter, Fabian},
title = {The Rise of the Galactic Empire: Ultraviolet Luminosity Functions at z $\sim$ 17 and z $\sim$ 25 Estimated with the MIDIS+NGDEEP Ultra-deep JWST/NIRCam Data Set},
journal = {The Astrophysical Journal},
}

@article{Castellano:2025vkt,
    author = "Castellano, M. and others",
    title = "{Pushing JWST to the extremes: Search and scrutiny of bright galaxy candidates at z {\ensuremath{\simeq}} 15{\textendash}30}",
    eprint = "2504.05893",
    archivePrefix = "arXiv",
    primaryClass = "astro-ph.GA",
    doi = "10.1051/0004-6361/202555082",
    journal = "Astron. Astrophys.",
    volume = "704",
    pages = "A158",
    year = "2025"
}

@article{Urrutia:2025fvp,
    author = "Urrutia, Juan and Ellis, John and Fairbairn, Malcolm and Vaskonen, Ville",
    title = "{Starlight from JWST: Implications for star formation and dark matter models}",
    eprint = "2504.20043",
    archivePrefix = "arXiv",
    primaryClass = "astro-ph.CO",
    reportNumber = "KCL-PH-TH/2025-12, CERN-TH-2025-085, AION-REPORT/2025-03",
    doi = "10.1051/0004-6361/202555390",
    journal = "Astron. Astrophys.",
    volume = "702",
    pages = "A109",
    year = "2025"
}

@article{Gorghetto:2025uls,
    author = "Gorghetto, Marco and Trifinopoulos, Sokratis and Valogiannis, Georgios",
    title = "{Large-Scale Structure Probes of the Post-Inflationary Axiverse}",
    eprint = "2511.04734",
    archivePrefix = "arXiv",
    primaryClass = "astro-ph.CO",
    reportNumber = "DESY-25-147, CERN-TH-2025-193",
    month = "11",
    year = "2025",
    journal="arxiv"   
}

@article{Ivanov:2025pbu,
    author = "Ivanov, Mikhail M. and Trifinopoulos, Sokratis",
    title = "{Effective Field Theory Constraints on Primordial Black Holes from the High-Redshift Lyman-{\ensuremath{\alpha}} Forest}",
    eprint = "2508.04767",
    archivePrefix = "arXiv",
    primaryClass = "astro-ph.CO",
    reportNumber = "MIT-CTP/5895, CERN-TH-2025-155",
    doi = "10.1103/8g8z-bmxd",
    journal = "Phys. Rev. Lett.",
    volume = "136",
    number = "17",
    pages = "171402",
    year = "2026"
}

@article{Rosenberg:2022smo,
    author = "Rosenberg, Erik and Gratton, Steven and Efstathiou, George",
    title = "{CMB power spectra and cosmological parameters from Planck PR4 with CamSpec}",
    eprint = "2205.10869",
    archivePrefix = "arXiv",
    primaryClass = "astro-ph.CO",
    journal = "Mon. Not. Roy. Astron. Soc.",
    volume = "517",
    number = "3",
    pages = "4620--4636",
    year = "2022"
}

@article{Efstathiou:2019mdh,
    author = "Efstathiou, George and Gratton, Steven",
    title = "{A Detailed Description of the CamSpec Likelihood Pipeline and a Reanalysis of the Planck High Frequency Maps}",
    eprint = "1910.00483",
    archivePrefix = "arXiv",
    primaryClass = "astro-ph.CO",
    journal = "Open J. Astrophys.",
    volume = "4",
    pages = "8",
    year = "2021"
}

@article{Afshordi:2003zb,
    author = "Afshordi, N. and McDonald, P. and Spergel, D. N.",
    title = "{Primordial black holes as dark matter: The Power spectrum and evaporation of early structures}",
    eprint = "astro-ph/0302035",
    archivePrefix = "arXiv",
    doi = "10.1086/378763",
    journal = "Astrophys. J. Lett.",
    volume = "594",
    pages = "L71--L74",
    year = "2003"
}

@article{Bucher:1999re,
    author = "Bucher, Martin and Moodley, Kavilan and Turok, Neil",
    title = "{The General primordial cosmic perturbation}",
    eprint = "astro-ph/9904231",
    archivePrefix = "arXiv",
    reportNumber = "DAMTP-1999-50",
    doi = "10.1103/PhysRevD.62.083508",
    journal = "Phys. Rev. D",
    volume = "62",
    pages = "083508",
    year = "2000"
}

@article{Elor:2023xbz,
    author = "Elor, Gilly and Jinno, Ryusuke and Kumar, Soubhik and McGehee, Robert and Tsai, Yuhsin",
    title = "{Finite Bubble Statistics Constrain Late Cosmological Phase Transitions}",
    eprint = "2311.16222",
    archivePrefix = "arXiv",
    primaryClass = "hep-ph",
    reportNumber = "FTPI-MINN-23-20, UTWI-39-2023",
    doi = "10.1103/PhysRevLett.133.211003",
    journal = "Phys. Rev. Lett.",
    volume = "133",
    number = "21",
    pages = "211003",
    year = "2024"
}

@article{Chung:2015tha,
      author         = "Chung, Daniel J. H.",
      title          = "{Large blue isocurvature spectral index signals
                        time-dependent mass}",
      journal        = "Phys. Rev.",
      volume         = "D94",
      year           = "2016",
      number         = "4",
      pages          = "043524",
      doi            = "10.1103/PhysRevD.94.043524",
      eprint         = "1509.05850",
      archivePrefix  = "arXiv",
      primaryClass   = "astro-ph.CO",
      SLACcitation   = "%%CITATION = ARXIV:1509.05850;%%"
}

@article{Kasuya:2009up,
      author         = "Kasuya, Shinta and Kawasaki, Masahiro",
      title          = "{Axion isocurvature fluctuations with extremely blue
                        spectrum}",
      journal        = "Phys.Rev.",
      volume         = "D80",
      pages          = "023516",
      doi            = "10.1103/PhysRevD.80.023516",
      year           = "2009",
      eprint         = "0904.3800",
      archivePrefix  = "arXiv",
      primaryClass   = "astro-ph.CO",
      SLACcitation   = "%%CITATION = ARXIV:0904.3800;%%",
}

@article{Planck:2018jri,
    author = "Akrami, Y. and others",
    collaboration = "Planck",
    title = "{Planck 2018 results. X. Constraints on inflation}",
    eprint = "1807.06211",
    archivePrefix = "arXiv",
    primaryClass = "astro-ph.CO",
    doi = "10.1051/0004-6361/201833887",
    journal = "Astron. Astrophys.",
    volume = "641",
    pages = "A10",
    year = "2020"
}

@article{ACT:2025tim,
    author = "Calabrese, Erminia and others",
    collaboration = "ACT",
    title = "{The Atacama Cosmology Telescope: DR6 Constraints on Extended Cosmological Models}",
    eprint = "2503.14454",
    archivePrefix = "arXiv",
    primaryClass = "astro-ph.CO",
    reportNumber = "FERMILAB-PUB-25-0157-PPD",
    month = "3",
    year = "2025"
}

@article{Pagano:2019tci,
    author = "Pagano, L. and Delouis, J. -M. and Mottet, S. and Puget, J. -L. and Vibert, L.",
    title = "{Reionization optical depth determination from Planck HFI data with ten percent accuracy}",
    eprint = "1908.09856",
    archivePrefix = "arXiv",
    primaryClass = "astro-ph.CO",
    doi = "10.1051/0004-6361/201936630",
    journal = "Astron. Astrophys.",
    volume = "635",
    pages = "A99",
    year = "2020"
}

@article{ACT:2023dou,
    author = "Qu, Frank J. and others",
    collaboration = "ACT",
    title = "{The Atacama Cosmology Telescope: A Measurement of the DR6 CMB Lensing Power Spectrum and Its Implications for Structure Growth}",
    eprint = "2304.05202",
    archivePrefix = "arXiv",
    primaryClass = "astro-ph.CO",
    reportNumber = "FERMILAB-PUB-23-237-PPD, FERMILAB-PUB-23-237-PPD",
    doi = "10.3847/1538-4357/acfe06",
    journal = "Astrophys. J.",
    volume = "962",
    number = "2",
    pages = "112",
    year = "2024"
}

@article{ACT:2023kun,
    author = "Madhavacheril, Mathew S. and others",
    collaboration = "ACT",
    title = "{The Atacama Cosmology Telescope: DR6 Gravitational Lensing Map and Cosmological Parameters}",
    eprint = "2304.05203",
    archivePrefix = "arXiv",
    primaryClass = "astro-ph.CO",
    reportNumber = "FERMILAB-PUB-23-206-PPD",
    doi = "10.3847/1538-4357/acff5f",
    journal = "Astrophys. J.",
    volume = "962",
    number = "2",
    pages = "113",
    year = "2024"
}

@article{Carron:2022eyg,
    author = "Carron, Julien and Mirmelstein, Mark and Lewis, Antony",
    title = "{CMB lensing from Planck PR4~maps}",
    eprint = "2206.07773",
    archivePrefix = "arXiv",
    primaryClass = "astro-ph.CO",
    doi = "10.1088/1475-7516/2022/09/039",
    journal = "JCAP",
    volume = "09",
    pages = "039",
    year = "2022"
}

@article{DESI:2024uvr,
    author = "Adame, A. G. and others",
    collaboration = "DESI",
    title = "{DESI 2024 III: baryon acoustic oscillations from galaxies and quasars}",
    eprint = "2404.03000",
    archivePrefix = "arXiv",
    primaryClass = "astro-ph.CO",
    reportNumber = "FERMILAB-PUB-24-0159-PPD",
    doi = "10.1088/1475-7516/2025/04/012",
    journal = "JCAP",
    volume = "04",
    pages = "012",
    year = "2025"
}

@article{DESI:2024lzq,
    author = "Adame, A. G. and others",
    collaboration = "DESI",
    title = "{DESI 2024 IV: Baryon Acoustic Oscillations from the Lyman alpha forest}",
    eprint = "2404.03001",
    archivePrefix = "arXiv",
    primaryClass = "astro-ph.CO",
    reportNumber = "FERMILAB-PUB-24-0147-PPD",
    doi = "10.1088/1475-7516/2025/01/124",
    journal = "JCAP",
    volume = "01",
    pages = "124",
    year = "2025"
}

@article{DESI:2024mwx,
    author = "Adame, A. G. and others",
    collaboration = "DESI",
    title = "{DESI 2024 VI: cosmological constraints from the measurements of baryon acoustic oscillations}",
    eprint = "2404.03002",
    archivePrefix = "arXiv",
    primaryClass = "astro-ph.CO",
    reportNumber = "FERMILAB-PUB-24-0154-PPD",
    doi = "10.1088/1475-7516/2025/02/021",
    journal = "JCAP",
    volume = "02",
    pages = "021",
    year = "2025"
}

@article{Torrado:2020dgo,
    author = "Torrado, Jesus and Lewis, Antony",
    title = "{Cobaya: Code for Bayesian Analysis of hierarchical physical models}",
    eprint = "2005.05290",
    archivePrefix = "arXiv",
    primaryClass = "astro-ph.IM",
    reportNumber = "TTK-20-15",
    doi = "10.1088/1475-7516/2021/05/057",
    journal = "JCAP",
    volume = "05",
    pages = "057",
    year = "2021"
}

@article{Chung:2024ctx,
    author = "Chung, Daniel J. H. and Tadepalli, Sai Chaitanya",
    title = "{Large blue spectral index from a conformal limit of a rotating complex~scalar}",
    eprint = "2406.12976",
    archivePrefix = "arXiv",
    primaryClass = "astro-ph.CO",
    doi = "10.1103/PhysRevD.111.083527",
    journal = "Phys. Rev. D",
    volume = "111",
    number = "8",
    pages = "083527",
    year = "2025"
}

@article{Blas:2011rf,
    author = "Blas, Diego and Lesgourgues, Julien and Tram, Thomas",
    title = "{The Cosmic Linear Anisotropy Solving System (CLASS) II: Approximation schemes}",
    eprint = "1104.2933",
    archivePrefix = "arXiv",
    primaryClass = "astro-ph.CO",
    reportNumber = "CERN-PH-TH-2011-082, LAPTH-010-11",
    doi = "10.1088/1475-7516/2011/07/034",
    journal = "JCAP",
    volume = "07",
    pages = "034",
    year = "2011"
}

@article{Chung:2023syw,
    author = {Chung, Daniel J. H. and M{\"u}nchmeyer, Moritz and Tadepalli, Sai Chaitanya},
    title = "{Search for isocurvature with large-scale structure: A forecast for Euclid and MegaMapper using EFTofLSS}",
    eprint = "2306.09456",
    archivePrefix = "arXiv",
    primaryClass = "astro-ph.CO",
    doi = "10.1103/PhysRevD.108.103542",
    journal = "Phys. Rev. D",
    volume = "108",
    number = "10",
    pages = "103542",
    year = "2023"
}

@article{Enander:2017ogx,
    author = "Enander, Jonas and Pargner, Andreas and Schwetz, Thomas",
    title = "{Axion minicluster power spectrum and mass function}",
    eprint = "1708.04466",
    archivePrefix = "arXiv",
    primaryClass = "astro-ph.CO",
    doi = "10.1088/1475-7516/2017/12/038",
    journal = "JCAP",
    volume = "12",
    pages = "038",
    year = "2017"
}

@article{Planck:2018lkk,
    author = "Aghanim, N. and others",
    collaboration = "Planck",
    title = "{Planck 2018 results. III. High Frequency Instrument data processing and frequency maps}",
    eprint = "1807.06207",
    archivePrefix = "arXiv",
    primaryClass = "astro-ph.CO",
    doi = "10.1051/0004-6361/201832909",
    journal = "Astron. Astrophys.",
    volume = "641",
    pages = "A3",
    year = "2020"
}

@article{Esteban:2023xpk,
    author = "Esteban, Ivan and Peter, Annika H. G. and Kim, Stacy Y.",
    title = "{Milky~Way satellite velocities reveal the dark matter power spectrum at small scales}",
    eprint = "2306.04674",
    archivePrefix = "arXiv",
    primaryClass = "astro-ph.CO",
    doi = "10.1103/PhysRevD.110.123013",
    journal = "Phys. Rev. D",
    volume = "110",
    number = "12",
    pages = "123013",
    year = "2024"
}

@article{Chung:2021lfg,
    author = "Chung, Daniel J. H. and Tadepalli, Sai Chaitanya",
    title = "{Analytic treatment of underdamped axionic blue isocurvature perturbations}",
    eprint = "2110.02272",
    archivePrefix = "arXiv",
    primaryClass = "astro-ph.CO",
    doi = "10.1103/PhysRevD.105.123511",
    journal = "Phys. Rev. D",
    volume = "105",
    number = "12",
    pages = "123511",
    year = "2022"
}

@article{Planck:2018vyg,
    author = "Aghanim, N. and others",
    collaboration = "Planck",
    title = "{Planck 2018 results. VI. Cosmological parameters}",
    eprint = "1807.06209",
    archivePrefix = "arXiv",
    primaryClass = "astro-ph.CO",
    doi = "10.1051/0004-6361/201833910",
    journal = "Astron. Astrophys.",
    volume = "641",
    pages = "A6",
    year = "2020",
    note = "[Erratum: Astron.Astrophys. 652, C4 (2021)]"
}

@article{Ghosh:2021axu,
    author = "Ghosh, Subhajit and Kumar, Soubhik and Tsai, Yuhsin",
    title = "{Free-streaming and coupled dark radiation isocurvature perturbations: constraints and application to the Hubble tension}",
    eprint = "2107.09076",
    archivePrefix = "arXiv",
    primaryClass = "astro-ph.CO",
    doi = "10.1088/1475-7516/2022/05/014",
    journal = "JCAP",
    volume = "05",
    number = "05",
    pages = "014",
    year = "2022"
}

@article{HERA:2022wmy,
    author = "Abdurashidova, Zara and others",
    collaboration = "HERA",
    title = "{Improved Constraints on the 21 cm EoR Power Spectrum and the X-Ray Heating of the IGM with HERA Phase I Observations}",
    eprint = "2210.04912",
    archivePrefix = "arXiv",
    primaryClass = "astro-ph.CO",
    doi = "10.3847/1538-4357/acaf50",
    journal = "Astrophys. J.",
    volume = "945",
    number = "2",
    pages = "124",
    year = "2023"
}

@article{deKruijf:2024voc,
    author = "de Kruijf, Jessie and Vanzan, Eleonora and Boddy, Kimberly K. and Raccanelli, Alvise and Bartolo, Nicola",
    title = "{Searching for blue-tilted power spectra in the dark ages}",
    eprint = "2408.04991",
    archivePrefix = "arXiv",
    primaryClass = "astro-ph.CO",
    reportNumber = "UTWI-26-2024",
    doi = "10.1103/PhysRevD.111.063507",
    journal = "Phys. Rev. D",
    volume = "111",
    number = "6",
    pages = "063507",
    year = "2025"
}

@article{Garcia-Gallego:2026phh,
    author = "Garcia-Gallego, Olga and Ir{\v{s}}i{\v{c}}, Vid and Viel, Matteo and Haehnelt, Martin G. and Bolton, James S.",
    title = "{Post-inflationary axion constraints from the Lyman-alpha forest}",
    eprint = "2603.04401",
    archivePrefix = "arXiv",
    primaryClass = "astro-ph.CO",
    month = "3",
    year = "2026",
    journal="arxiv"
}

@article{Pavicevic:2025gqi,
    author = "Pavi{\v{c}}evi{\'c}, Mak and Ir{\v{s}}i{\v{c}}, Vid and Viel, Matteo and Bolton, James S. and Haehnelt, Martin G. and Martin-Alvarez, Sergio and Puchwein, Ewald and Ralegankar, Pranjal",
    title = "{Constraints on Primordial Magnetic Fields from the Lyman-alpha Forest}",
    eprint = "2501.06299",
    archivePrefix = "arXiv",
    primaryClass = "astro-ph.CO",
    doi = "10.1103/77rd-vkpz",
    journal = "Phys. Rev. Lett.",
    volume = "135",
    number = "7",
    pages = "071001",
    year = "2025"
}

\end{document}